\documentclass[floatfix,aps,prl,abstract,twocolumn,10pt,nofootinbib,preprintnumbers]{revtex4-1}

\usepackage{amsgen,amsmath,amstext,amsbsy,amsopn,amssymb}
\usepackage{graphicx}
\usepackage[usenames]{color}
\usepackage{epsfig}
\usepackage{multirow}
\usepackage{bm}
\usepackage[colorlinks=true, pdfstartview=FitV, linkcolor=blue, citecolor=cyan, urlcolor=magenta]{hyperref}
\usepackage{hyperref}
\usepackage{lineno}

\newcommand{\Ddel}{\delta_{\rm D}   }

\newcommand{\MpcOh}{ \,  \mathrm{Mpc}  \, h^{-1} }
\newcommand{\hOMpc}{ \,  \mathrm{Mpc}^{-1}  \, h  }

\newcommand{\comment}[1]{}
\newcommand{\nn}{ \nonumber }

\newcommand{\Msun}{ \,   M_{\odot}  {h}^{-1}   }

\newcommand{\beq}{\begin{equation}}
\newcommand{\eeq}{\end{equation}}
\newcommand{\beqa}{\begin{eqnarray}}
\newcommand{\eeqa}{\end{eqnarray}}

\usepackage[OT2,T1]{fontenc}
\DeclareSymbolFont{cyrletters}{OT2}{wncyr}{m}{n}
\DeclareMathSymbol{\Sha}{\mathalpha}{cyrletters}{"58}

\begin{document}

%\preprint{Version 0.1}

\title{Volume Statistics as a Probe of Large-Scale Structure}

\author{Kwan Chuen Chan$^{(1)}$ }
\email{chankc@mail.sysu.edu.cn}
\author{Nico Hamaus$^{(2)}$ }

\affiliation{$^{1}$ School of Physics and Astronomy, Sun Yat-sen University, 2 Daxue Road, Tangjia, Zhuhai, 519082, China }

\affiliation{ $^{2}$Universit\"ats-Sternwarte M\"unchen, Fakult\"at f\"ur Physik, Ludwig-Maximilians Universit\"at, Scheinerstr.~1, 81679 M\"unchen, Germany}

\date{\today}

\begin{abstract}
  We investigate the application of volume statistics to probe the distribution of underdense regions in the large-scale structure of the Universe. This statistic measures the distortion of Eulerian volume elements relative to Lagrangian ones and can be built from tracer particles using tessellation methods. We apply Voronoi and Delaunay tessellation to study the clustering properties of density and volume statistics. Their level of shot-noise contamination is similar, as both methods take into account all available tracer particles in the field estimator. The tessellation causes a smoothing effect in the power spectrum, which can be approximated by a constant window function on large scales. The clustering bias of the volume statistic with respect to the dark matter density field is determined and found to be negative. We further identify the Baryon Acoustic Oscillation (BAO) feature in the volume statistic. Apart from being smoothed out on small scales, the BAO is present in the volume power spectrum as well, without any systematic bias. These observations suggest that the exploitation of volume statistics as a complementary probe of cosmology is very promising.
\end{abstract}

\maketitle

\section{Introduction}

Modern large-scale structure surveys probe both the background expansion and the growth of structure, which can be used to constrain the energy content of the Universe and help resolve the mystery behind the cosmic expansion \cite{WeinbergMortonson_etal2013}. These surveys have been remarkably successful in constraining the parameters in $\Lambda$CDM and its extensions \cite{Cole:2005sx,Eisenstein:2005su,Beutler2011_6dFGS,Blake2012_WiggleZ, Anderson_BOSS2012, Ross2015_MSG, Alam:2016hwk,Alam:2020sor}. Because most observed galaxies trace the high-density regions of the Universe \cite{Kaiser1984}, previous analyses typically focus on the clustering statistics of overdensities. However, the dominant part of the Universe exhibits a low density compared to its mean value. The clustering of underdensities may therefore contain additional information that is complementary to that obtained via conventional methods.

Cosmic voids are extended regions of relatively low matter content in the large-scale structure, they are potentially ideal proxies for studying the underdense parts of the Universe \cite{ WeygaertvanKampen1993,Gottloeber:2003zb,Sheth:2003py}. Because of their unique environment, interesting phenomena such as the dark energy \cite{PisaniSutter_etal2015,Pollina:2015uaa,Sahlen:2018cku,Verza:2019tvg}, modified gravity \cite{ClampittCaiLi_2013,Cai_etal2015, Zivick_etal2015, Achitouv:2016jjj, Perico:2019obq}, and the influence of massive neutrinos \cite{Massara:2015msa,Kreisch:2018var,Schuster:2019hyl,Contarini:2020fdu} become more visible in voids. Recently, there have been numerous studies on the clustering of voids using samples derived from cosmological surveys, exploiting effects such as Redshift-Space Distortions (RSD) in the void-galaxy correlation function \cite{Paz:2013sza, Hamaus:2015yza,Cai:2016jek,Achitouv:2016mbn,Hawken:2016qcy,Chuang:2016wqb,Hamaus:2017dwj,Aubert_etal2020,Nadathur:2020vld}, the Alcock-Paczynski test \cite{Lavaux:2009wm,Sutter_etal2012,Hamaus:2016wka,Mao:2016faj,Correa:2018vge,Hamaus:2020cbu}, and the Baryon Acoustic Oscillations (BAO) \cite{Liang:2015oqc,Kitaura:2015ubm}. However, the spatial number density of voids is generally low due to their large size; hence their auto-clustering statistic exhibits a considerable shot noise contribution~\cite{HamausWandeltetal2014, Sutter:2013ssy,Hamaus:2014afa}.  While small voids are more abundant, they actually reside in regions of higher density (so-called voids-in-clouds \cite{Sheth:2003py}), only larger voids truly trace the underdense regime of large-scale structure. To mitigate the shot noise, cross-correlations with other more abundant tracers, such as galaxies~\cite{HamausWandeltetal2014} or clusters \cite{Pollina:2016gsi,Pollina:2018ekp}, are considered. Another option is the use of overlapping spherical voids, as demonstrated by Ref.~\cite{Kitaura:2015ubm} in the measurement of the BAO in the void auto-correlation function. Allowing voids to overlap increases their number density by up to two orders of magnitude, but the signal-to-noise ratio of the resulting clustering statistic is more difficult to interpret.

In practice voids are identified via sparse tracers, as the full mass distribution cannot be observed directly in all three dimensions. Since many tracers are needed to define a single void, the resulting void distribution is even more sparse.  Therefore, the question arises whether one can construct a field estimator for the low-density regions with low level of shot noise and without overlap.  The {\it volume statistic} proposed in this paper satisfies these conditions. In Sec.~\ref{sec:vol_statistics}, we define the volume statistic in detail. In order to implement it in data, we need to estimate the density field using tessellation methods. We briefly review two tessellation interpolation methods and contrast them with the conventional mass assignment scheme in Sec.~\ref{sec:tessellation_interpolation}. The possibility of approximating its impact on the power spectrum by an effective window is discussed in Sec.~\ref{sec:tessellation_as_effectivewindow}. Sec.~\ref{sec:volume_clustering} is devoted to studying the overall clustering properties of the volume statistic, such as the bias parameter and the BAO feature. We conclude in Sec.~\ref{sec:conclusions}. In the appendices, we review the the Kernel Density Estimation method (Appendix~\ref{sec:Density_KDE}) and the effects of a window function on the power spectrum (Appendix~\ref{sec:window_Pk}).

\section{Volume statistics }
\label{sec:vol_statistics}

Voids are extended underdense regions in the Universe and it has been shown that they furnish a biased tracer of the large-scale structure \cite{Sheth:2003py,HamausWandeltetal2014,Chan:2014qka, Clampitt:2015jra, Chan:2018piq,Schuster:2019hyl,Chan:2019yzq,Jamieson:2019dmp}. Small voids tend to reside in overdense environments and therefore exhibit positive bias parameters; they are in the so-called void-in-cloud regime \cite{Sheth:2003py}. Large voids, on the contrary, are anti-correlated with overdense structures and thus exhibit a negative clustering bias \cite{HamausWandeltetal2014}, i.e. they trace large-scale underdensities (it is interesting to note that the quadratic bias of voids follows a similar trend to that of halos \cite{Chan:2019yzq}). However, their number density is significantly lower than that of small voids (owing to their size), which results in a high shot-noise contribution and hence a low signal-to-noise ratio of their large-scale auto-power spectrum. In this work, instead of defining single objects from a large number of tracers, we assign a volume to each individual tracer particle, which helps to mitigate the shot noise.

In the Lagrangian picture of large-scale structure, we can think of each tracer particle carrying a volume element that is being distorted over time. This evolution is governed by mass conservation, and for the case of dark matter we have~\cite{PTreview} 
\begin{align}
  \label{eq:mass_conservation_DM}
  d^3 q &=  ( 1 +\delta ) d^3 x, 
\end{align}
where  $ d^3q $ and  $d^3x $ are the volume elements in the Lagrangian (initial) and Eulerian (evolved) space, respectively, and $\delta$ is the matter density contrast in Eulerian space. Then we define
\beq
\mathcal{J}   \equiv   \frac{ d^3 x  }{ d^3 q } = \frac{1 }{1 + \delta }.  
\eeq
$\mathcal{J} $ is the Jacobian of the transformation from the Lagrangian coordinate $\bm{q} $ to the Eulerian coordinate $\bm{x} $. It quantifies the change in volume compared to that in the Lagrangian space. When the displacement is small, $ \mathcal{J} $ reduces to $ 1 - \delta $ with a symmetric distribution around a mean equal to 1. For large displacements and hence nonlinear $\delta $, $\mathcal{J}$ has a mean larger than 1 and the distribution becomes skew symmetric with a singular tail at $\delta = -1$. This indicates that inside empty regions in Eulerian space, a tremendous amount of expansion in volume has occurred.  After shell crossing, the volume element $d^3q $ can split into multiple parts and the fluid description by Eq.~\eqref{eq:mass_conservation_DM} is no longer valid. Nonetheless, shell crossing occurs mainly in the high-density regions, which are down-weighted in $\mathcal{J}$, and hence its impact on our statistic is minimal.

In analogy to the density contrast, we can define the {\it volume statistic} as
\beq
\label{eq:Delta}
\mathcal{V} \equiv \frac{  \mathcal{J}  }{ \bar{  \mathcal{J}  } } - 1 = \frac{1}{ \bar{  \mathcal{J}  } ( 1 + \delta ) } - 1, 
\eeq
where $  \bar{  \mathcal{J}} $ is the mean of $ \mathcal{J} $. For small $\delta$, it reduces to $ - \delta $.

For halos the situation is more complicated, because the density is no longer uniform in Lagrangian space, and we have instead  (e.g.~\cite{Matsubara:2008wx,Matsubara2011IPT})
\begin{align}
  \label{eq:delta_h_continuity}
(1 + \delta_{\rm L} )d^3 q & =   (1 + \delta_{\rm h} )  d^3 x ,  
%    \frac{ d^3 x   }{  d^3 q } &= \frac{ 1 + \delta_{\rm L} }{ 1 + \delta }. 
\end{align}
where  $\delta_{\rm L}$ and $\delta_{\rm h} $ are the halo density contrast in Lagrangian and Eulerian space, respectively.   However, in observations we have no direct access to $\delta_{\rm L} $. One practical way to bypass this difficulty is to define  $ \mathcal{J} $ as
\begin{align}
\mathcal{J} & \equiv \frac{d^3 x  }{ ( 1 + \delta_{\rm L} )  d^3 q  } = \frac{1 }{  1 + \delta_{\rm h} } 
\end{align}
even for the case of halos.  We can think of it as the volume change inversely weighted by the halo density in Lagrangian space.

The advantage of the volume statistic is as follows. First, it is a fundamental measure of the volume change induced by the large-scale structure evolution. From a theoretical perspective it is likely more amenable to models than the statistics of voids, which are defined by the complicated nonlinear topology of the cosmic web. In the linear regime the volume statistic reduces to the one of linear density fluctuations. In the weakly nonlinear regime it can be treated with perturbation theory or other analytic methods.  Second, the volume statistic makes use of all the available tracer particles, resulting in a shot-noise level comparable to its original tracer field.  Nevertheless, because the majority of particles are located inside collapsed regions, the sampling is still poor compared to the conventional density statistic. On the other hand, because of the denominator $1 + \delta $, $\mathcal{V} $ is undefined when $\delta = -1 $. This can pose problems in the density estimation for sparse samples, such as massive halos. We will therefore consider tessellation methods to avoid such singularities.

The volume statistic up-weights underdense regions and down-weights overdense ones; hence it is similar to certain types of marked correlation functions in spirit\footnote{This can be generalized to the marked power spectrum \cite{Massara:2020pli,Philcox:2020srd}.}. If each particle is assigned a mark $m$, the marked correlation $M$ for the mark can be generally defined as \cite{Sheth:2005aj}
\beq
M(r) = \frac{ 1 }{ N(r) \bar{m}^2 } \sum_{ij} m_i m_j,  % = \frac{ 1+ \mathcal{W} }{ 1 + \xi },  
\eeq
where $N(r)$ is the total number of pairs with separation $r$ and  $\bar{m} $ is the mean of the mark.  The mark can be chosen to highlight any particular physics of interest. For example, \cite{White:2016yhs} proposed to consider a mark of the form
\beq
m = \left( \frac{  \rho_{ * } + 1  }{   \rho_{ * } + 1  + \delta   } \right)^p, 
\eeq
where $ \rho_{ * } $ and $p$ are free parameters. This mark has been applied to study modified gravity \cite{Valogiannis:2017yxm}. If we set $ \rho_{ * } = 0 $ and $p =1$, it reduces to the volume statistic. However, the value of $ \rho_{ * }  $ is often taken to be order unity, e.g.~in~\cite{Valogiannis:2017yxm}, so as to avoid the singularity problem mentioned above.

Moreover, the definition of the volume statistic is similar to the log-transform of the density field \cite{Neyrinck:2009fs}
\beq
A = \log ( 1 + \delta ). 
\eeq
It was found that the field $A$ is more Gaussian than $\delta $ at late times, and some higher-order information in $\delta$ is pulled back into the two-point statistics of $A$~\cite{Neyrinck:2009fs,Wang:2011fj}. Both transformations up-weight the low-density regions and down-weight the high-density ones. In fact, both the log-field and the volume statistic diverge at $\delta = -1 $. The log-transform aims to recover the information in the initial density field, as it accommodates both the low-density and the high-density regions. Our goal is less ambitious and we merely want to recover the information from the underdense regions. By restricting our target, it is easier to define a density reconstruction method that suits precisely this purpose.

\section{Interpolation methods}
\label{sec:tessellation_interpolation}

In order to compute the volume statistic one needs a density estimate that is finite everywhere. To this end we use interpolation by tessellation to reconstruct the density field from a discrete particle distribution. As this type of method is rarely used directly for clustering analyses, we shall first compare it against the commonly used mass interpolation method.

\subsection{Mass interpolation}

For the large-scale clustering analysis, the density field is often interpolated to a grid\footnote{The configuration-space estimation of correlation functions uses counting of pairs directly without the need for a grid.}. The interpolation methods usually employed include Nearest Grid Point (NGP), Cloud-in-Cell (CIC), and Triangular-Shaped Cloud (TSC). We collectively denote them as mass-interpolation methods, they are discussed extensively in the monograph \cite{HockneyEastwood1981}. In these methods, particles are smoothed by a window function (or cloud function) $W$ of fixed size and these windows can overlap with each other. The density at the grid point $x_p$ is given by
\beq
\rho ( \bm{x}_p) =  \sum_\alpha  \int_{ [ \bm{x}_p ] } d^3 x \,  W( \bm{x} -  \bm{x}_\alpha ), 
\eeq
where $ [ \bm{x}_p ] $ denotes  the grid cell enclosing $ \bm{x}_p $ and the sum is over all the particles.

The three-dimensional (3D) window can be constructed from its 1D form as $W(\bm{x} ) = W(x)W(y)W(z)$.  In 1D, the NGP window function is the Dirac delta, for CIC it is a top-hat, and for TSC it is a triangular-shaped cloud centered on the particle position. In fact these are the first three members of an infinite hierarchy of interpolating functions that can be generated via iterative convolution with the top-hat window function~\cite{HockneyEastwood1981}. In NGP the particle mass is interpolated to the nearest grid point, in CIC the mass is assigned to the grid points of the cell enclosing the particle (8 points in 3D), and the particle mass is interpolated to the neighboring 27 points in TSC. Thus, the window becomes more extended as the order of the interpolation scheme increases. Because the window function is a constant linear operator, its effect can be removed by division of $W$ in Fourier space. To represent the density field for clustering analyses, the grid size is usually chosen sufficiently large to mitigate resolution effects\footnote{Discrete sampling of the continuous field introduces aliasing effect \cite{HockneyEastwood1981,Jing:2004fq,Sefusatti:2015aex}, which is especially serious for modes close to the Nyquist frequency. We are limited to the large-scale modes because of the tessellation interpolation, we shall not consider it here.}. Among the mass-interpolation methods, CIC is the most common one, so we use it as our benchmark.

\subsection{Tessellation interpolation}

The tessellation method can generate a space-filling field with finite values (non-vanishing) everywhere in configuration space; it is often used as an intermediate step to obtain a smooth density field and to define extended structures in the cosmic web, e.g.~\cite{ Neyrinck_etal_VOBOZ,Neyrinck:2007gy,Sousbie2011,Cautun_etal2013Nexus,Cautun_etal2014Nexus,Busch:2019gwv,Paranjape:2020wuc}. However, it has rarely been used directly for clustering analyses of the density field.  Given a set of points (generators) $\bm{x}_\alpha $,  a tessellation generates cells that are space-filling and mutually disjoint \cite{Press:2007:NRE:1403886,vandeWeygaert:2007ze}. Applications of the tessellation methods in cosmology were introduced in \cite{IckevandeWeygaert1987,YoshiokaIkeuchi1989ApJ, vandeWeygaert1994, BernardeauvandeWeygaert1996,Schaap:2000se,vandeWeygaert:2007ze}.

In the Voronoi tessellation, each Voronoi cell encloses one generator and every point $\bm{x}$ belongs to a Voronoi cell $V_\alpha$ if the distance $ | \bm{x} -  \bm{x}_\alpha | $ is smaller than  $| \bm{x} -  \bm{x}_\beta | $ for any other $\beta$. The simplest possibility is to assume that the particle mass spreads out uniformly within a Voronoi cell and the density within each cell is estimated by the mass of the particle and the volume of the Voronoi cell.  This method yields a piece-wise constant field and the field value is discontinuous across the boundaries. We shall refer to the technique to estimate densities in this way as the Voronoi Tessellation Field Estimator (VTFE).

To obtain a continuous field we can tessellate the space into Delaunay cells. In 3D, a triangulation divides space into tetrahedrons. A Delaunay tessellation is such a triangulation with the additional property that the circumsphere of each tetrahedron does not contain any generators.   The field value inside a tetrahedron can then be obtained by linear interpolation from the vertices.   Explicitly, the number density at a point $\bm{x}$ inside a Delaunay cell with vertices $\bm{x}_0$, $\bm{x}_1$, $\bm{x}_2$, and $\bm{x}_3$, is given by~\cite{vandeWeygaert:2007ze} 
\begin{align}
  \label{eq:DTFE_linear_intp}
n(\bm{x} ) = n(\bm{x}_0 ) + ( \bm{x} - \bm{x}_0  )_i J^{-1}_{\quad i \alpha } \Delta n_\alpha,    
\end{align}
where $ \Delta n_\alpha $ indicates the difference in density between vertex $ \bm{x}_\alpha $ and the reference vertex $ \bm{x}_0 $,   $ n( \bm{x}_\alpha )  -  n( \bm{x}_0 ) $. A Latin index denotes the spatial component of the vector and a Greek index represents the vertex (also from 1 to 3). The Jacobian $J$ is the transformation matrix between the coordinate $\bm{x}$ and the barycentric coordinate~$\zeta_\alpha $  
\begin{align}
    J_{i\alpha}=\frac{ \partial \bm{x}_i }{ \partial  \bm{\zeta}_\alpha }  \approx  ( \bm{x}_{\alpha}  - \bm{x}_0 )_i  .
\end{align}
The number density at the generator can be estimated by \cite{Schaap:2000se,vandeWeygaert:2007ze} 
\beq
n( \bm{x}_\alpha ) =   \frac{ 1 + D  }{ \sum_{ \mathcal{D}_{  \bm{x}_\alpha  }^{\rm adj}}  \mathcal{V}_{\rm adj} (  \bm{x}_\alpha ) }, 
\eeq
where $D$ is the dimension of space, $\mathcal{D}_{  \bm{x}_\alpha  }^{\rm adj} $ denotes all the adjacent Delaunay cells with $ \bm{x}_\alpha $ as one of their generators, and $ \mathcal{V}_{\rm adj}  $ is the volume of the adjacent Delaunay cells. This construction ensures that the integral of the  density field obtained from Eq.~\eqref{eq:DTFE_linear_intp} is equal to the total particle mass.  We refer to this method of estimating the density as the Delaunay Tessellation Field Estimator (DTFE) in this paper.

Effectively, tessellation methods smooth out the particle mass adaptively based on the local density. For high-density regions, the effective window size is small, while it is large in the low-density regime. For the mass-interpolation methods, such as the CIC, as the size of the grid increases the resulting density field approaches a sum of Dirac delta distributions. Thus, this method can produce the same type of distribution as the point set. On the other hand, the tessellation method aims at reconstructing a smooth (piecewise constant for VTFE and continuous for DTFE) density field from the discrete particle distribution. This additional requirement introduces a smoothing to the small-scale power spectrum. The smoothing scale is dictated by the mean particle separation of the distribution. In summary, in the mass-interpolation scheme, the effective window-function size is controlled by the grid size, while that of the tessellation method is determined by the mean particle separation.

\subsection{Effects of the tessellation: An analytic estimate}
\label{sec:tessellation_KDE}

In the mass-interpolation scheme, a fixed window is used to smooth the particle distribution. An analytic estimate can be made if the particles are assumed to Poisson sample the underlying density field\footnote{In the case of halos and voids there are exclusion effects that violate this assumption \cite{Hamaus:2010im,Baldauf:2013hka,HamausWandeltetal2014,Chan:2014qka}}. The density estimated by smoothing the particle distribution  is akin to the kernel density estimation (KDE) method in statistics, which is used to estimate the probability density distribution (see e.g.~\cite{LiRacine_KDE}). In Appendix \ref{sec:Density_KDE}, we review the KDE and show that given a sample of particles it can be used to estimate the number density of the sample, $ n $ [Eq.~\eqref{eq:n_KDEestimator}]. In particular, it is shown that smoothing introduces a bias to the density estimate [Eq.~\eqref{eq:KDEn_bias}]. At a peak (trough) of the density field, the window-estimated density field is biased low (high).  However, the advantage of smoothing the density field is that it can suppress the variance of the estimator [Eq.~\eqref{eq:KDEn_var}]: it is inversely proportional to the average volume of the window size. In statistics, the mean squared error, which is a sum of the variance and the bias squared, matters if the signal-to-noise ratio of the measurement is not high. Thus the KDE is useful if its suppression to the variance is larger than its bias.

For the case of density estimation by tessellation, the density at the position $\bm{x}$ depends on the particles spanning the cell. It is hard to make an analytic estimate of its bias and variance, because of the irregular shape of the cell and the interpolation scheme depending on all the points in the cell. For the case of VTFE, because there is only one generator inside a Voronoi cell, we can write down the density estimate as  ($N=1$)
\begin{align}
\hat{n}_{\rm VTFE} (\bm{x}) = \frac{1  }{h_\alpha^3 } W_{\rm VTFE}  \left(  \frac{\bm{x} - \bm{x}_\alpha  }{ h_\alpha }  \right), 
\end{align}
where $ W_{\rm VTFE} $ describes the Voronoi cell around a point $\bm{x}$, $\bm{x}_\alpha $ is the generator of the Voronoi cell, and $h_\alpha $ schematically represents the characteristic size of the cell.

Because the Voronoi cell is generally not symmetric about its generator, the bias reads 
\begin{align}
  \label{eq:VTFE_bias}
 \langle \hat{n}(\bm{x} ) \rangle  -  n(\bm{x}) = - h_\alpha \sum_i I_i^{(1)} \frac{\partial n }{\partial x_i } +          \frac{  h_{\alpha}^2  }{2}  \sum_{ij} \frac{\partial^2 n }{ \partial x_i \partial x_j  }  I^{(1)}_{ij}    ,   
\end{align}
where the function $I$ is defined in general in Eq.~\eqref{eq:I_integral}. In particular, because of the irregular shape of the Voronoi cells,  $I^{(1)} $ is in general non-vanishing.   However, if we now further average over the shape of the Voronoi cell $W_{\rm VTFE} $, $\langle  I_i^{(1)} \rangle $ vanishes and we obtain
\begin{align}
  \label{eq:VTFE_bias_2}
 \langle \hat{n}(\bm{x} ) \rangle  -  n(\bm{x}) =    \frac{ \langle  h_{\alpha}^2 \rangle  }{2}  \sum_{ij} \frac{\partial^2 n }{ \partial x_i \partial x_j  } \langle  I^{(1)}_{ij} \rangle .    
\end{align}
Similar to the standard KDE case [Eq.~\eqref{eq:KDEn_bias}], we recover the result that the bias is sourced by the second derivatives of the density. The variance of the estimator reads 
\begin{align}
  \mathrm{Var}( \hat{n}(\bm{x} ) ) &=\frac{1}{h_\alpha^6 } \mathrm{Var} \left( W_{\rm VTFE}  \Big(\frac{\bm{x} - \bm{x}_\alpha }{ h_\alpha } \Big) \right) \nn \\
  & \approx  \frac{ I^{(2)} }{ h_\alpha^3 } n(\bm{x} ). 
\end{align}
Further averaging over the shape of the Voronoi cell, we have
\begin{align}
 \mathrm{Var}( \hat{n}(\bm{x} ) ) &= \left\langle \frac{1 }{ h_\alpha^3 } \right\rangle \langle  I^{(2)} \rangle  n(\bm{x} ).
\end{align}

For DTFE, the density at $\bm{x} $ [Eq.~\eqref{eq:DTFE_linear_intp}] depends on the position of the vertices nonlinearly and the previous analytic arguments do not apply. Nonetheless, we shall see that the features of the DTFE are qualitatively similar to those of the VTFE.

\section{Effective window function}
\label{sec:tessellation_as_effectivewindow}

In this section we investigate whether the effects of the tessellation can be modeled by an effective window function. The width of the window involved in the mass interpolation is controlled by the grid size, which can be chosen such that the grid effects are negligible, or the window function effect can be divided out in Fourier space. Hence we may treat the density field obtained from the CIC, $\delta_{\rm CIC} $, as the unsmoothed field for comparison.

In Appendix~\ref{sec:window_Pk} we review the effects of the window function on the power spectrum.  We show that if the particle distribution is smoothed by a constant window function $W$, the effect on the power spectrum is captured by an overall factor of $|W|^2$ (or $W$ for the case of cross-power spectrum). Note that the same window function factor applies to both the continuous clustering and the shot-noise term. Thus, if the effective window function can be taken to be a non-stochastic function, the power spectrum of the density field obtained with the tessellation method, $ \delta_{\rm tess}  $, can be written as\footnote{The Fourier convention used in this paper is \\$ f(\bm{k}) =  \int  \frac{d^3 x   }{(2 \pi)^3  } e^{- i \bm{k} \cdot \bm{x} } f(\bm{x})  $ and $ f(\bm{x}) =  \int  d^3 k  e^{ i \bm{k} \cdot \bm{x} } f(\bm{k})  $.}
\beq
\label{eq:Weff_a}
\langle \delta_{\rm tess}(\bm{k}) \delta_{\rm tess}(\bm{k}') \rangle = (2 \pi)^6  | W_{\rm tess} (\bm{k}) |^2   \langle \delta_{\rm CIC} (\bm{k})   \delta_{\rm CIC} (\bm{k}') \rangle.    
\eeq
Similarly for consistency, the cross-correlation is given by 
\beq
\label{eq:Weff_c}
\langle \delta_{\rm tess}(\bm{k}) \delta_{\rm CIC}(\bm{k}') \rangle =  (2 \pi)^3  W_{\rm tess}(\bm{k})   \langle \delta_{\rm CIC} (\bm{k})   \delta_{\rm CIC} (\bm{k}') \rangle.    
\eeq

We test these relations using simulation data, the details of which are described in the following. The assumed cosmology is a flat $\Lambda$CDM model with parameters $\Omega_{\rm m}=0.3  $,  $\Omega_{\Lambda}=0.7  $, $h_0 =0.7$, $n_{\rm s} = 0.967$, and  $\sigma_8=0.85 $. Each simulation consists of a cubic box of $ 1000 \MpcOh $ side length and contains $ 512^3 $ particles. The Gaussian initial conditions are generated by {\tt CLASS} \cite{CLASS_code} at redshift $z=49$. In order to study the BAO signature, we ran another set of simulations with Einsenstein-Hu initial conditions without the BAO wiggles \cite{EisensteinHu_1998} (NoWiggle). The particle displacements are implemented using {\tt 2LPTic}~\citep{Crocce:2006ve} and  are evolved with the $N$-body code {\tt Gadget2} \citep{Springel:2005mi}. Halos are identified using the halo finder {\tt AHF}~\citep{Knollmann:2009pb}. They are defined via a spherical overdensity threshold of 200 times the background density and contain at least 20 particles. For each simulation set, 20 realizations are used. The results derived from the halo samples will be our primary interest in this paper. We shall illustrate our results using four halo samples at $z=0$ and $1$, with two sets of initial conditions, respectively. Unless otherwise stated, the error bars are the standard error of the mean of the measurements among all realizations.

The procedures for density estimation using the tessellation method  are similar to those in the public code {\tt DTFE } \cite{CautunvdWeygaert2011}, which implements the density estimation by the Delaunay tessellation efficiently.  We implement both Voronoi and Delaunay tessellation methods using the python libraries based on the {\tt Qhull} code \cite{QHull}. After generating the tessellation cells we use a cubic grid to estimate the density at the grid points, which can then be used for the estimation of power spectra. We use a grid size of $256^3$ and average over $2^3$ uniformly spaced sampling points within each grid cell to calculate the mean density at the grid point. We have checked that these choices are sufficiently accurate for our results.

\subsection{Measurement of the effective window}
\label{sec:window_tessellation}

We measure the effective window function of the tessellation using random catalogs and halo samples.   Without intrinsic clustering it is easier to identify the impact of the effective window on the power spectrum, since the remaining clustering amplitude is simply white noise. To generate the random catalogs, particle positions are randomly placed in a cubic box of size $1000 \MpcOh $ until the desired number density is reached (we match it with the halo sample density). Assuming that halos Poisson sample the underlying continuous density field, the shot-noise contribution to their power spectrum is given by~\cite{Peebles}
\beq
\label{eq:Pshot}
P_{\rm  shot} = \frac{1}{  (2 \pi)^3 \bar{n} },
\eeq
where $ \bar{n} $ is the mean halo density. We show in Appendix~\ref{sec:window_Pk} that there is an additional factor of $|W|^2$, if the particle distribution is smoothed by a window $W$.

\begin{figure}[!tb]
\centering
\includegraphics[width=\linewidth]{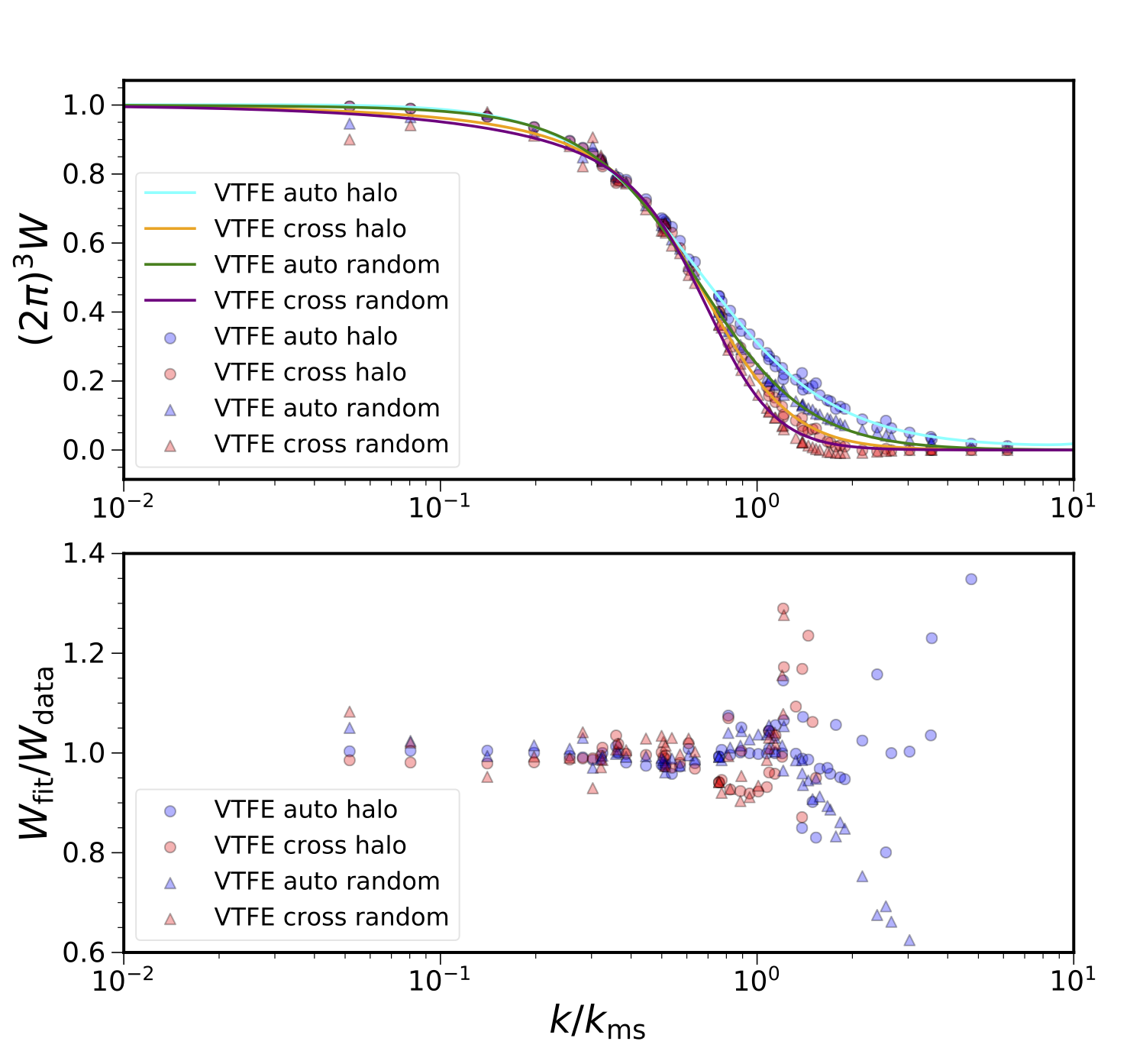}
\caption{Top: the VTFE effective window function measured from the random catalog (triangles) and the halos (circles). Measurements from the auto-power spectrum (blue) and the cross-power spectrum (red) are shown. The fits using Eq.~\eqref{eq:Weff_fit} are over-plotted (cyan for $W_{\rm a}$ from halos, green for $W_{\rm a}$  from randoms, orange for $W_{\rm c}$ from halos and purple for $W_{\rm c}  $ from randoms). Bottom: the ratio between the fits and the measurements (circles for halos and triangles for randoms, blue for auto- and red for cross-power spectra).}
\label{fig:Weff_random_data_fit_VTFE}
\end{figure}

The effective VTFE window obtained using Eqs.~\eqref{eq:Weff_a} and \eqref{eq:Weff_c} from the random catalog is shown in Fig.~\ref{fig:Weff_random_data_fit_VTFE}. In this case the CIC power spectrum is given by Eq.~\eqref{eq:Pshot}. The size of the effective window can be characterized by the mean separation wavenumber $  k_{\rm ms} \equiv  \pi \bar{n}^{1/3}  $. We find that using the variable $q \equiv k / k_{\rm ms}$, the effective window obtained from different tracers falls on a universal curve.  Thus, $k_{\rm ms}$ represents the characteristic size of the tessellation cell in Fourier space\footnote{We note that in Lagrangian space, the typical extension of halo density profiles is controlled by halo mass. When the density profile is rescaled by this characteristic extension, most of the mass dependence is removed and the halo density profiles fall on a rather universal curve  \cite{ChanShethScoccimarro2017}. Thus, $k_{\rm ms}$ is analogous to the role of mass for Lagrangian halos.}.  Although the window from the auto- and cross-power spectra (denoted as $W_{\rm a}$ and  $W_{\rm c}$, respectively) agree with each other at low $k$, $W_{\rm a} $ is more extended than $W_{\rm c} $. In fact, $W_{\rm c}$  turns negative at $q \sim  1.8 $, while $W_{\rm a}$  remains positive everywhere. Eqs.~\eqref{eq:Weff_a} and \eqref{eq:Weff_c} imply that we can extract the same window function from the auto- and cross-power spectrum measurements. However, in practice, this only works on large scales, so the assumption that $W(\bm{k})$ factorizes and does not explicitly depend on the density field $\delta(\bm{k})$ must fail on smaller scales.

%If the window function is really a constant function, Eqs.~\eqref{eq:Weff_a} and \eqref{eq:Weff_c} imply that we would get the same window function from the auto and cross measurements. Since this only holds on large scales, we conlcude that taking the window to be a non-stochastic function is a valid assumption only on large scales and this is consistent with the interpretation that the effective window is density dependent.

Even if the particle distribution is intrinsically clustered, the window function still factors out. In Fig.~\ref{fig:Weff_random_data_fit_VTFE} we also show the effective window function extracted from $ P_{\rm tess} / P_{\rm CIC} $ and  $ P_{\rm tess,CIC} / P_{\rm CIC} $, with $ P_{\rm tess} $ and $ P_{\rm CIC} $ being the halo auto-power spectra obtained with the tessellation and the CIC method, and $  P_{\rm tess,CIC} $ the cross-power spectra between them. Note that we have not subtracted shot noise in any of those cases. The curves from halo samples are somewhat less universal than for the randoms, especially at high $q$. This is expected, as the halos exhibit intrinsic clustering that is scale dependent, making their effective window more extended due to stronger clustering on smaller scales.

We find that the overall window function shape can be described well by the functional form 
\begin{align}
  \label{eq:Weff_fit}
 (2\pi)^3  W(q) =\frac{ 1 }{ ( 1 + a q )[ 1 +(bq)^n ] }, 
\end{align}
with $q = k / k_{\rm ms} $.

\begin{figure}[!tb]
\centering
\includegraphics[width=0.98\linewidth]{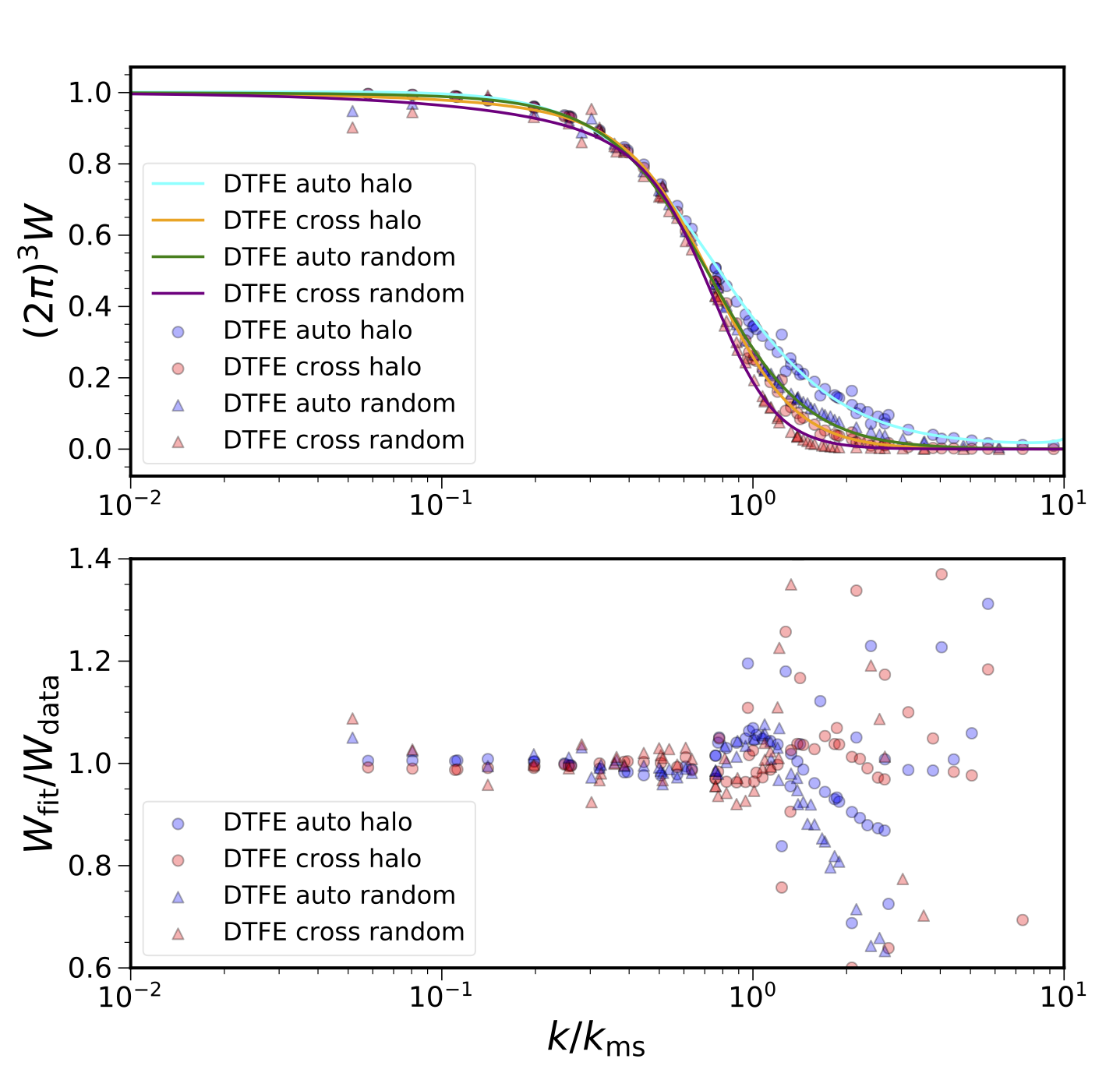}
\caption{As Fig.~\ref{fig:Weff_random_data_fit_VTFE}, but for the DTFE results. } 
\label{fig:Weff_random_data_fit_DTFE}
\end{figure}

We have shown the best-fit values obtained from the VTFE and the DTFE in Table \ref{tab:fitting_paramters}. The results for the halo and random samples obtained with $W_a$ and $W_c$ are compared.   These best fits are also plotted in Figs.~\ref{fig:Weff_random_data_fit_VTFE} and \ref{fig:Weff_random_data_fit_DTFE}.  Up to $ q \lesssim 1 $, the accuracy of the fit is about~$10\%$ within the scatter of the data. As there are significant differences between $W_{\rm a}$ and $W_{\rm c}$, obtaining a universal fit that is accurate at $ q \gtrsim 1 $ is not possible. Overall, the DTFE results are in qualitative agreement with the VTFE ones. However, we find that the measurements of $ W_{\rm DTFE} $ are slightly more extended than  $ W_{\rm VTFE} $. This implies that the DTFE window is more compact in configuration space.

%For the halo sample the best-fit values are  $ a = -0.08 $, $ b=1.55 $, and $ n=2.08 $ for $W_a$ ($ a = 0.37 $, $ b=1.35 $, and $ n=3.14 $ for $W_{\rm c}$). The corresponding results for the random catalog  are $ a = 0.09 $, $ b=1.49 $, and $ n=2.47 $  for $W_a$ ($ a = 0.50 $, $ b=1.36 $, and $ n=3.94 $ for $W_{\rm c}$). These best fits are also shown in Fig.~\ref{fig:Weff_random_data_fit_VTFE}. Up to $ q \lesssim 1 $, the accuracy of the fit is about~$10\%$ within the scatter of the data. As there are significant differences between $W_{\rm a}$ and $W_{\rm c}$, obtaining a universal fit that is accurate at $ q \gtrsim 1 $ is not possible.

%The corresponding results for the DTFE are presented in Fig.~\ref{fig:Weff_random_data_fit_DTFE}. Overall, they are qualitatively in agreement with the VTFE case. However, we find that the measurements of $ W_{\rm DTFE} $ are slightly more extended than  $ W_{\rm VTFE} $. This implies that the DTFE window is more compact in configuration space. We have also fitted the DTFE window using Eq.~\eqref{eq:Weff_fit} and the results are:  for the halo sample, $ a = -0.09 $, $ b=1.37 $, and $ n=2.20 $ for  $W_a$  ($ a = 0.20 $, $ b=1.28 $, and $ n=3.25 $ for $W_{\rm c}$).  For the random sample, $ a = 0.07 $, $ b=1.36 $, and $ n=2.75 $ for $W_a$  ($ a = 0.37 $, $ b=1.29 $, and $ n=4.18 $ for $W_{\rm c}$). The quality of the fits is similar to the case of VTFE. 

\begin{table}
\centering
\caption{ The best-fit parameters for Eq.~\eqref{eq:Weff_fit} for the VTFE and DTFE methods.  The results derived from $W_a$  and $ W_c$ for the random and halo samples are shown.  }
\label{tab:fitting_paramters}
\begin{tabular}{|l|cc|cc|cc|cc|} 
\hline
\hline
 \multirow{2}{*}{} &
      \multicolumn{2}{c|}{VTFE random} &
      \multicolumn{2}{c|}{VTFE halo}  &
      \multicolumn{2}{c|}{DTFE random}  &
      \multicolumn{2}{c|}{DTFE halo}   \\
      \hline 
	& $W_a$ & $W_c$ & $W_a$ &  $ W_c $   &  $W_a$ & $W_c$ & $W_a$ &  $ W_c $ \\  
\hline
 $a$   &   0.09  &   0.50   &   -0.08  &  0.37  &  0.07  &  0.37  & -0.09   & 0.20   \\         
 $b$   &  1.49   &  1.36    &  1.55    & 1.35   &  1.36  &  1.29  &  1.37   & 1.28    \\
 $n$   &  2.47   &  3.94    &  2.08    &  3.14  &  2.75  &  4.18  &  2.20   & 3.25    \\
\hline
\hline
\end{tabular}
\end{table}

\subsection{Power spectrum ratio}

\begin{figure}[!tb]
\centering
\includegraphics[width=0.98\linewidth]{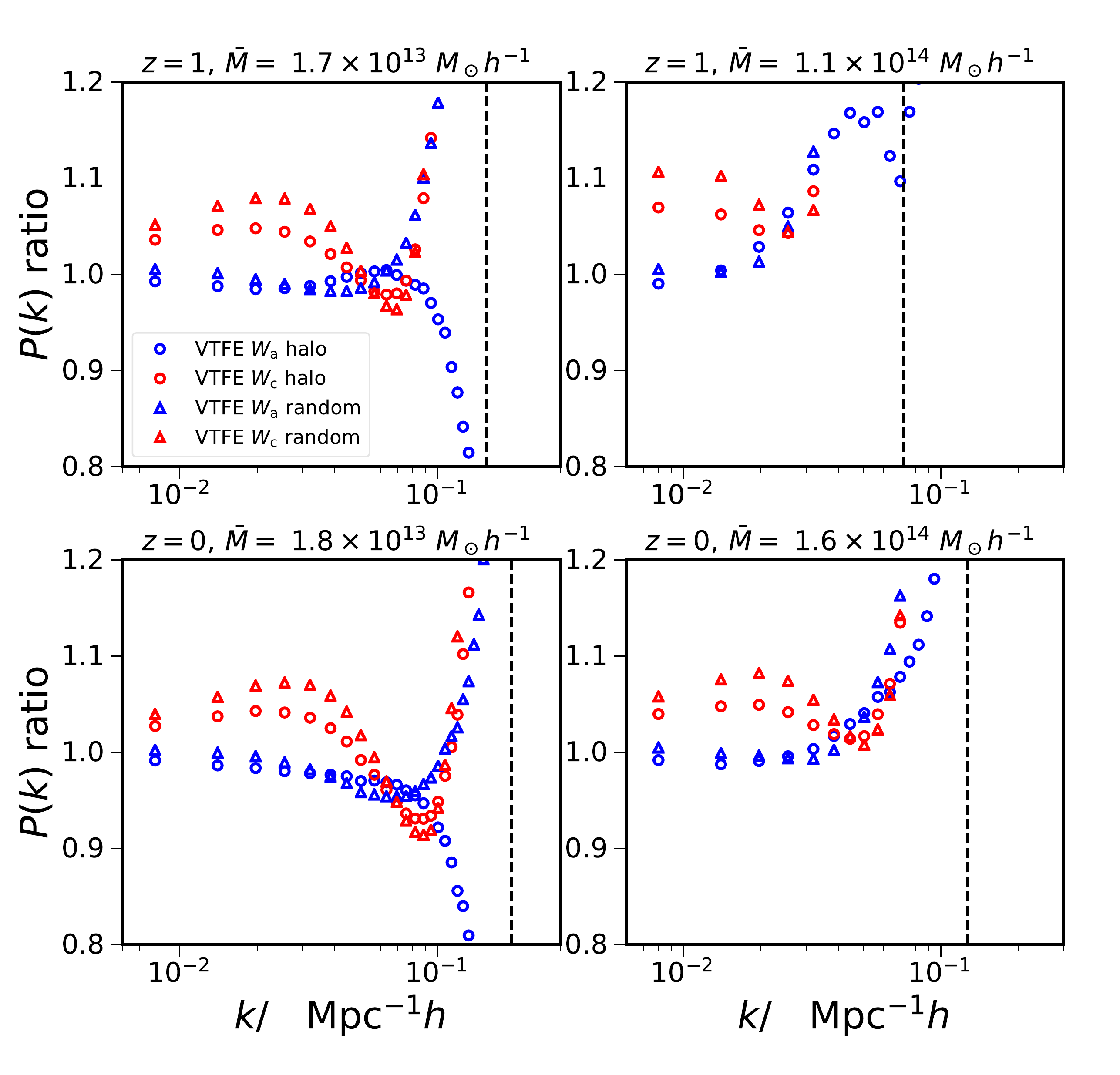}
\caption{Auto-power spectrum ratio Eq.~\eqref{eq:Pk_ratio_check} from the VTFE window for $W_{\rm a } $ (blue) and $W_{\rm c}$ (red) from the halo (circles) and random (triangles) samples. The mean mass and redshift of the halo samples are shown on top of each panel.  } 
\label{fig:Pkratio_Weff_division_VTFE_Pauto}
\end{figure}

\begin{figure}[!tb]
\centering
\includegraphics[width=\linewidth]{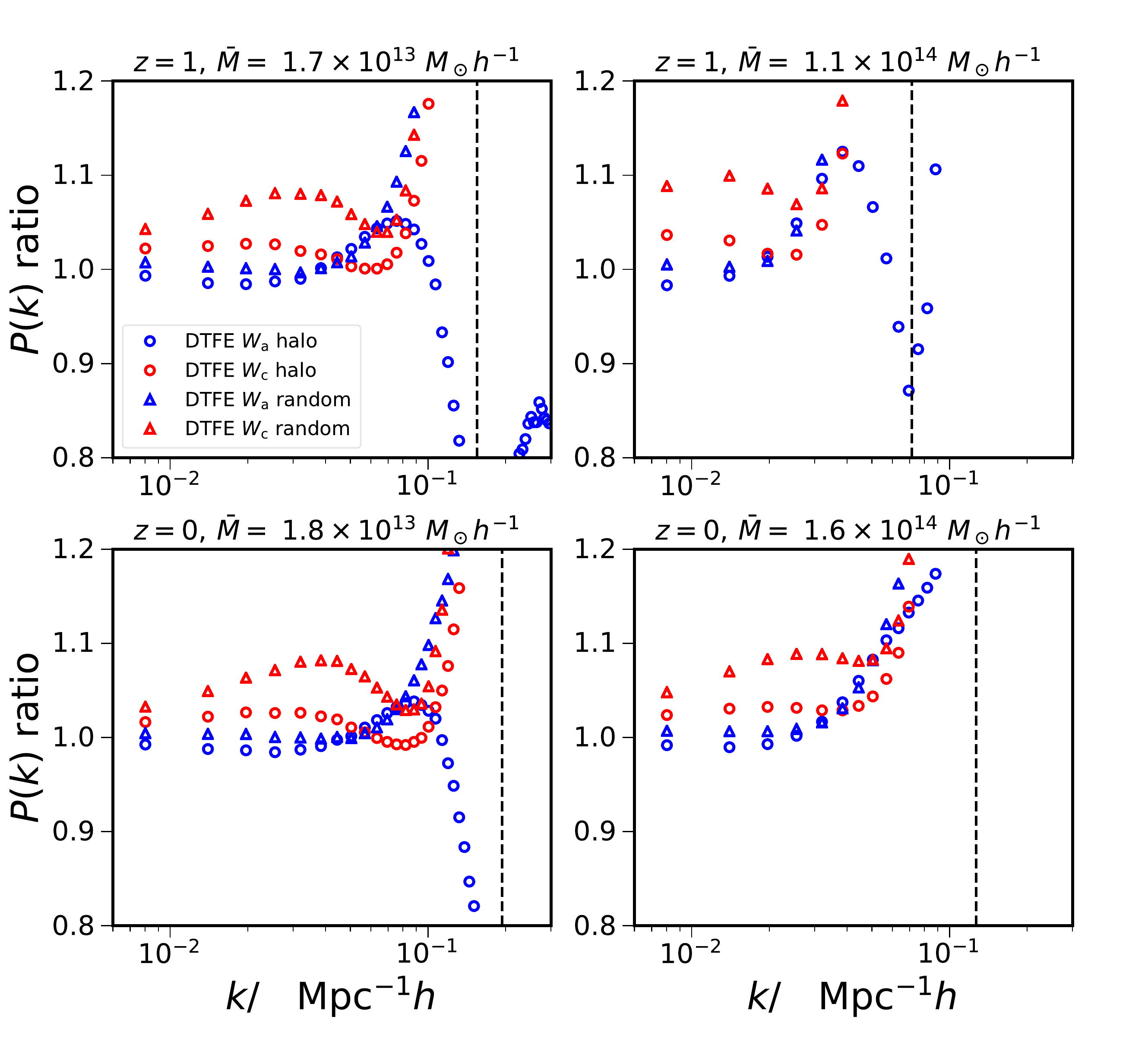}
\caption{As Fig.~\ref{fig:Pkratio_Weff_division_VTFE_Pauto}, but for the DTFE results. }
\label{fig:Pkratio_Weff_division_DTFE_Pauto}
\end{figure}

After having measured the effective window functions of the tessellation methods, we are now in the position to test whether we can use this window to recover the unsmoothed power spectrum. We utilize the ratio
\beq
\label{eq:Pk_ratio_check}
R = \frac{    \frac{ P_{\rm tess}^{\rm X} }{ [(2 \pi)^3 W(k) ]^n } - P_{\rm shot} }{  P_{\rm CIC} -  P_{\rm shot}  } ,
\eeq
to quantify the accuracy of this reconstruction, where $\mathrm{X}$ denotes auto- ($n = 2$) or cross-power spectra ($n=1$).  In Fig.~\ref{fig:Pkratio_Weff_division_VTFE_Pauto} we show the ratio $R$ for the auto-power spectrum using the VTFE windows $W_{\rm a}$ and $W_{\rm c}$ from halo and random samples. For the rest of this paper, we will exclusively show the results based on these four samples.  At low $k$, $W_{\rm a} $ appears to yield a better agreement than $W_{\rm c} $. However, inspection of Fig.~\ref{fig:Weff_random_data_fit_VTFE} reveals that this can be explained by the fact that $W_{\rm c}$ is a bit lower than $W_{\rm a}$ at small $q$. This is an artifact of the fitting formula.  On the other hand, when $k$ is close to $k_{\rm ms}$, the ratio deviates substantially from 1. A division by the window function is prone to noise and systematic error, which become inflated at high $k$ when $W$ is small. Even if we use $W_{\rm c}$ from a direct measurement, the pattern at high $k$ in $R$ remains unchanged. Also, the trends are very similar for both the halo and the random samples. In Fig.~\ref{fig:Pkratio_Weff_division_DTFE_Pauto} we show the corresponding results for the DTFE, which are qualitatively similar to the VTFE case. The ratios using the cross-power spectra in Eq.~\eqref{eq:Pk_ratio_check} are very similar, so we do not show them here.

Our results suggest that approximating the tessellation with a constant effective window function is valid on large scales, but above $k_{\rm ms }$ the density-dependent nature of the tessellation method is important and smoothing effects cannot be removed by division of a window function. In this case, it is necessary to determine the tessellation-smoothed density field numerically.

\begin{figure*}%[!htb]
\centering
\includegraphics[width=0.8\linewidth]{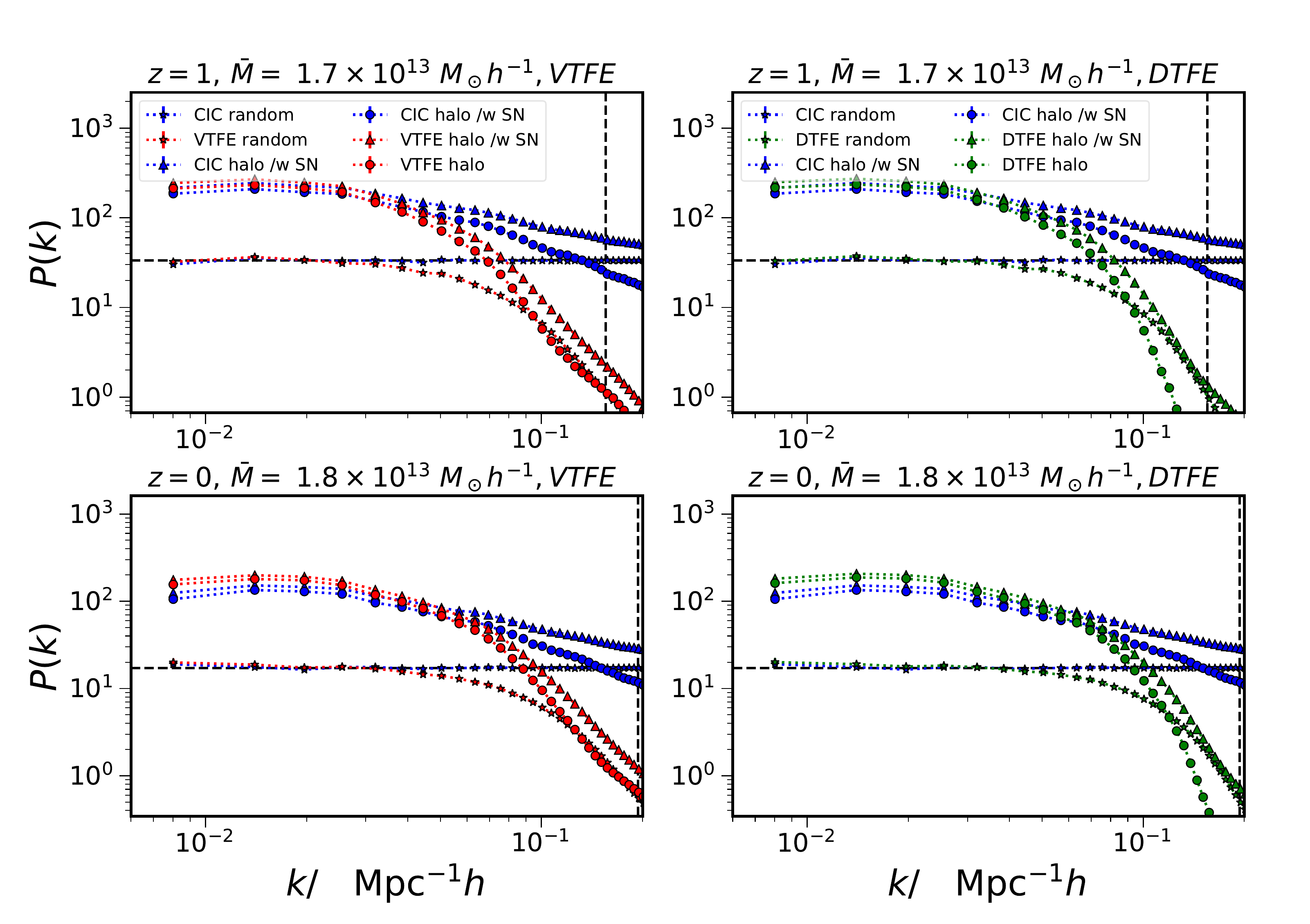}
\caption{Auto-power spectra of the volume statistic before (triangles) and after shot-noise subtraction (circles) for the VTFE (left panels, red) and the DTFE (right panels, green). The conventional halo-density power spectra are shown for comparison (blue), along with the shot noise contamination estimated from the randoms (stars). The vertical dashed line indicates the mean tracer separation scale $k_{\rm ms}$ and the horizontal one shows the Poisson power spectrum $ P_{\rm shot} $. The mean mass and redshift of the halo samples are shown on top of each panel. On large scales, the shot-noise level of the volume field is comparable to that of the halo sample used to construct it.   }
\label{fig:Pk_Vol_Shot_fullPk_VTFE_DTFE_split}
\end{figure*}

\begin{figure*}%[!htb]
    \centering
    \begin{minipage}{.75\linewidth}
    %\begin{subfigure}{.5\linewidth}
        \centering
        \includegraphics[width=\linewidth]{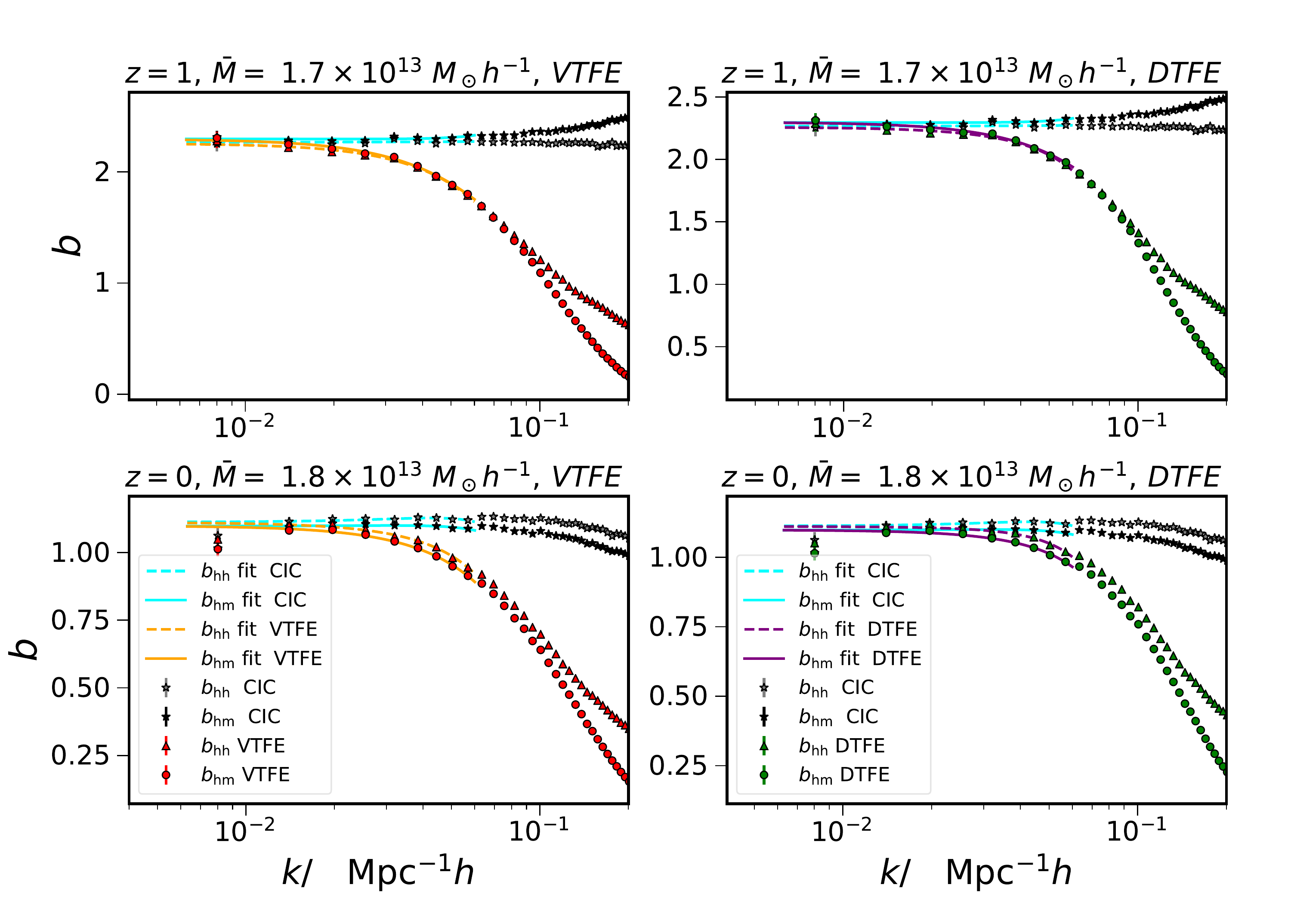}
        \caption{ The clustering bias of the halo density field obtained with the CIC, VTFE (left panels), and DTFE (right panels) methods. The measurements originate from the halo auto-power spectrum, $ b_{\rm hh} $ (CIC: grey stars, VTFE: red triangles, DTFE: green triangles) and the halo-matter cross-power spectrum, $ b_{\rm hm} $ (CIC: black stars, VTFE: red circles, DTFE: green circles). The fit using Eq.~\eqref{eq:b_fit} is over-plotted, dashed for  $ b_{\rm hh} $ and solid for $ b_{\rm hm} $; cyan, orange, and purple for CIC, VTFE, and DTFE measurements, respectively. Redshifts and mean halo masses are indicated on top of each panel. On large scales, the bias measurements obtained with the tessellation methods are in good agreement with the CIC ones.  }
        \label{fig:bias_estimator_flag1_subset}
    \end{minipage}%
    
        %\end{subfigure}%
    \begin{minipage}{0.75\linewidth}
    %\begin{subfigure}{.5\linewidth}
        \centering
        \includegraphics[width=\linewidth]{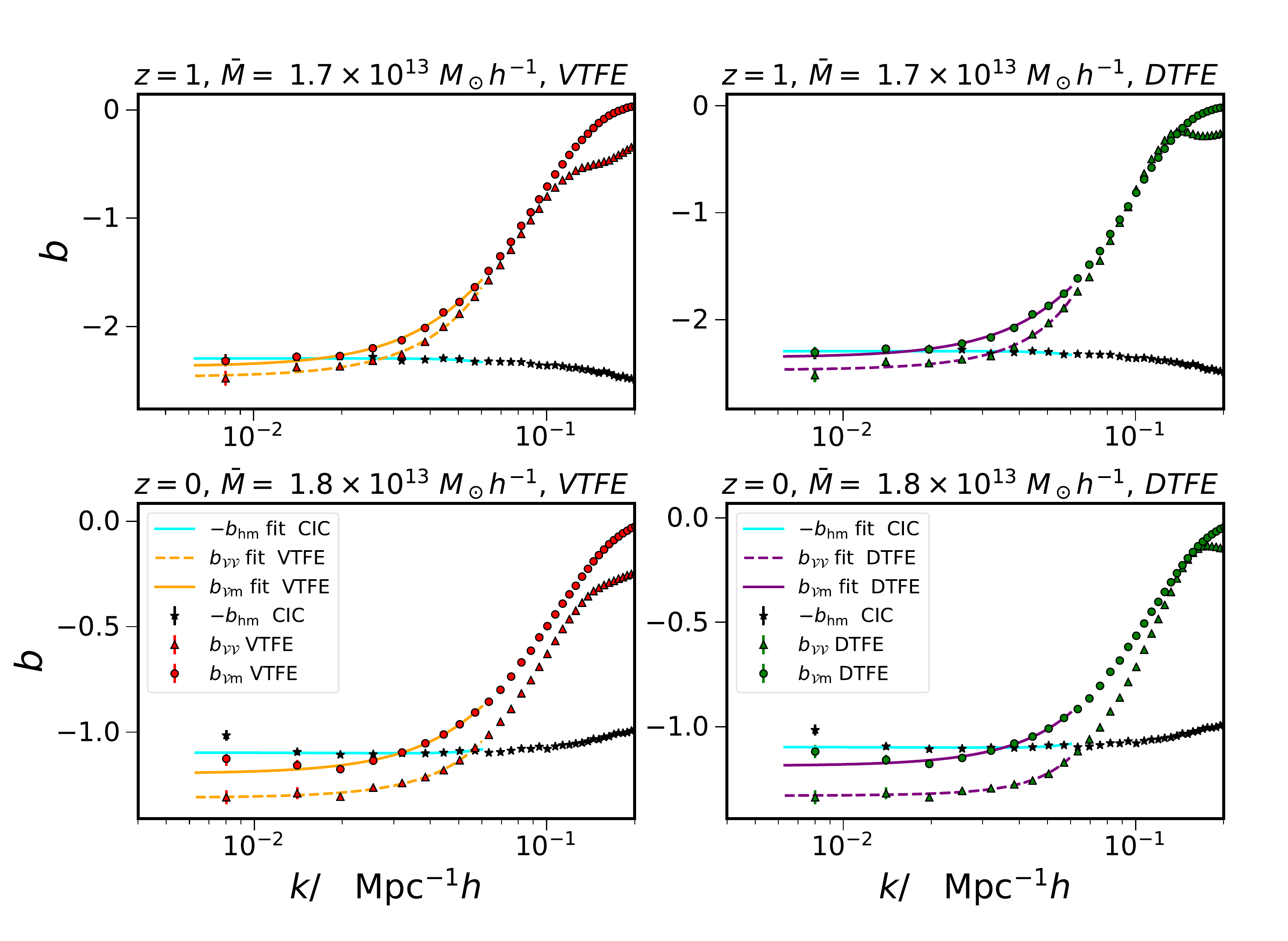}
        \caption{ As Fig.~\ref{fig:bias_estimator_flag1_subset}, but for the clustering bias of the volume field $\mathcal{V}$. To compare with the magnitude of the conventional halo-density bias, the negative of $b_{\rm hm}$ is shown. The properties of the halo samples used to construct the volume statistics are indicated on top of each panel. The bias of the volume field is negative and its magnitude is similar to that of the halo sample.  }
        \label{fig:bias_estimator_flag2_subset}
    \end{minipage}
    %\end{subfigure}%
\end{figure*}

\section{Clustering of the volume field}
\label{sec:volume_clustering}

\subsection{The shot noise of the volume field}

We estimate the shot noise of the volume field  $ P_{ \mathcal{V}  \rm n} $ using the random catalog again and showcase the auto-power spectra of the volume statistic before and after shot noise subtraction in Fig.~\ref{fig:Pk_Vol_Shot_fullPk_VTFE_DTFE_split}.   Although the volume statistic is obtained via nonlinear transformation of the halo tracer field, on large scales the shot noise still approaches $P_{\rm shot } $ as given by Eq.~\eqref{eq:Pshot}. However, it is suppressed by the effects of the tessellation on smaller scales. After shot noise is subtracted from the power spectra, the results from the VTFE and DTFE agree with the shot-noise subtracted CIC case. Thus, since all available tracer particles are used to construct the volume field, its shot-noise contamination is indeed similar to that of the original tracer field.

\subsection{The bias of the volume field }

The volume field can be regarded as a biased tracer of the underlying dark matter density field. As it traces the underdense regions of large-scale structure, we expect it to exhibit a negative clustering bias, similar to large voids \cite{Sheth:2003py,HamausWandeltetal2014,Chan:2014qka}. To guide our interpretation of the volume statistic, we first investigate the bias of the more common density field obtained via the tessellation method in Fig.~\ref{fig:bias_estimator_flag1_subset}. We compare the bias parameters $b_{\rm hh} $ and $b_{\rm hm} $ derived from the halo auto- and  halo-matter cross-power spectra, $P_{\rm hh} $ and $P_{\rm hm} $, 
\begin{align}
  \label{eq:bhh}
  b_{\rm hh} &= \sqrt{  \frac{ P_{\rm hh} - P_{ \rm hn} }{ P_{\rm mm} }   }, \\
  \label{eq:bhm}
  b_{\rm hm} &= \frac{ P_{\rm hm } }{ P_{\rm mm} },  
\end{align}
where $ P_{\rm mm} $ denotes the matter auto-power spectrum and $P_{ \rm hn}$ is the halo shot noise contribution obtained from the random catalog. As a comparison, we also show $b_{\rm hh}$ and $ b_{\rm hm}$ obtained from the CIC method.

On large scales the density-bias parameter obtained via the tessellation methods approaches the one obtained via the CIC mass assignment, which already reaches a constant value at $k \lesssim 0.08 \hOMpc$. On smaller scales, the bias parameter from the tessellation is damped by the effective window function, as discussed in the previous section. The differences between $b_{\rm hh} $ and $ b_{\rm hm}$  are apparent for $k \gtrsim 0.1 \hOMpc $. Consistent with the findings in Sec.~\ref{sec:window_tessellation}, Fig.~\ref{fig:bias_estimator_flag1_subset} shows that the DTFE window is more compact than the VTFE window in configuration space. To fit the large-scale bias function, the scale dependence of the window must be taken into account  and a possible functional form is a quartic polynomial 
\beq
\label{eq:b_fit}
b (k) = c_0 + c_2 k^2 + c_4 k^4 ,  
\eeq
where $c_0$, $c_2$, and  $c_4 $ are the fit parameters.  In Fig.~\ref{fig:bias_estimator_flag1_subset} we have also plotted the best-fit curves, using modes up to $k_{\rm max} =  0.06 \hOMpc$, yielding a good agreement with the simulation data.

We now turn to the bias of the volume field, presented in Fig.~\ref{fig:bias_estimator_flag2_subset}. The auto- and cross-bias for the volume field is defined in analogy to Eqs.~\eqref{eq:bhh} and \eqref{eq:bhm},
\begin{align}
  \label{eq:bVV}
  b_{ \mathcal{V} \mathcal{V} } &= - \sqrt{  \frac{ P_{  \mathcal{V} \mathcal{V} } - P_{ \mathcal{V}  \rm n} }{ P_{\rm mm} }   }, \\
  \label{eq:bVm}
  b_{\mathcal{V}  \rm m} &= \frac{ P_{ \mathcal{V}  \rm m } }{ P_{\rm mm} },  
\end{align}
with $ P_{  \mathcal{V} \mathcal{V} } $ being the auto-power spectrum of the volume field and $P_{ \mathcal{V}  \rm m } $ the cross-power spectrum between the volume field and the matter density field. The shot noise of the volume field  $ P_{ \mathcal{V}  \rm n} $ is estimated using the random catalog again. Note that because the volume field anti-correlates with the dark matter density field, its bias is {\it negative}. To ease comparison with the magnitude of the halo bias, we have also shown  $- b_{\rm hm} $. The overall shape of the bias functions of the volume field is similar to the density ones. On large scales, the bias of the volume field approaches the halo bias in magnitude. This is expected for large-scale fluctuations that are small, since in that limit $ \mathcal{V}$ reduces to  $- \delta_{\rm h} $.  However, although $b_{\mathcal{V}   \rm m}$ agrees with the halo bias well, there is a marked deviation of  $b_{ \mathcal{V}\mathcal{V}  }$ from the former. We presume there to be loop corrections to the power spectrum of quadratic order in density, analogous to the shot-noise renormalization effect in the local bias case (\cite{McDonald2006_biasrenorm}, see also \cite{Wang:2011fj}). These additional shot-noise-like contributions cause deviations from the linear bias on large scales. On small scales, the density field is suppressed by the tessellation and the definition of $\mathcal{V}  $  is designed such that it approaches zero in the limit of vanishing $\delta_{\rm h}$. Hence, the behavior of the bias of  $ \mathcal{V}$  is qualitatively similar to that of the tessellation density field.

Like cosmic voids, the volume field furnishes a negatively biased tracer of large-scale structure. Although the small-scale power is suppressed by the tessellation, the large-scale field is proportional to the underlying density field. In scenarios involving local primordial non-Gaussianity (PNG), the void bias exhibits a scale dependence on large linear scales \cite{Chan:2018piq}. Because the amplitude of void bias can be negative, this may be used to complement the traditional halo bias in constraining PNG~\cite{Dalal:2007cu}. Ref.~\cite{Chan:2018piq} demonstrated that a combination of halos and voids, taking advantage of the so-called multi-tracer approach~\cite{Seljak:2008xr,McDonald:2008sh}, allows to substantially tighten constraints on the non-Gaussianity parameter $f_{\rm NL}$. However, the gain in the constraining power of that analysis is limited by the shot noise in void auto-clustering statistics. As the volume field also exhibits a negative bias, similar gains can be expected, but with a lower level of shot-noise contamination.  One can think of the volume field as a ``dual'' of the density field with negative bias within the same survey volume. Thanks to their very different bias amplitudes, but comparable shot-noise levels, the density and volume fields together may provide optimal conditions for conducting a multi-tracer analysis \cite{HamausSeljakDesjacques2011, Hamaus:2012ap,Abramo:2013awa}.

\begin{figure*}%[!hbt]
\centering
\includegraphics[width=0.8\linewidth]{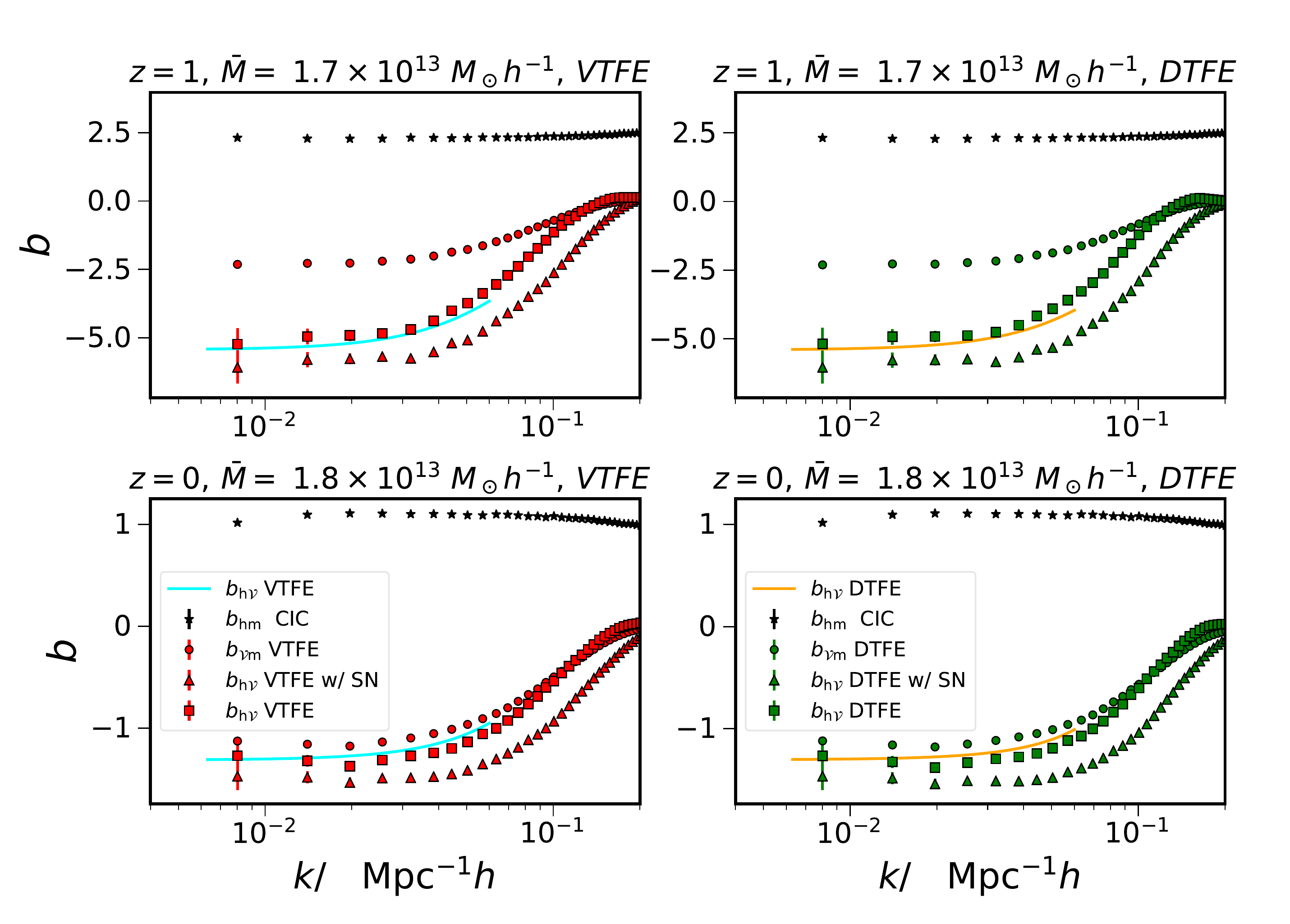}
\caption{The clustering bias $ b_{\rm h \mathcal{V} } $ obtained from the cross-power spectrum between the halo density field and the volume field using the VTFE (left panels, red) and the DTFE (right panels, green) methods. Because the volume field is derived from the halo distribution, there is a shot-noise contribution. Both the results before (triangles) and after shot-noise subtraction (squares) are compared. The halo bias $b_{\rm hm } $ (black stars) and volume bias $b_{ \mathcal{ V} \rm m } $  (VTFE: red circles, DTFE: green circles) are shown for reference. The solid curves (cyan for VTFE and orange for DTFE) are the predictions obtained using the fit results from Eq.~(\ref{eq:b_fit}), which are in good agreement with the direct measurements. }
\label{fig:bhV_bhmbVm_subset}
\end{figure*}

Although  $ b_{\mathcal{V} \rm m } $ is not directly observable in galaxy surveys, this can be circumvented by considering
\beq
 b_{\rm h \mathcal{V} } = \frac{ P_{ \rm h \mathcal{V} } }{P_{\rm mm} },
\eeq
where $ P_{ \rm h \mathcal{V} } $ is the observable cross-power spectrum between $ \delta_{\rm h }$ and $  \mathcal{V} $. The dark matter power spectrum $P_{\rm mm}$ can be modeled numerically or using perturbation theory. On large scales, where the fluctuations are small and the bias is linear, we have $  b_{\rm h \mathcal{V} } \approx b_{\rm h m} b_{ \mathcal{V}  \rm m} $. In Fig.~\ref{fig:bhV_bhmbVm_subset} we plot the   $ b_{\rm h \mathcal{V} } $ measurement from our simulation. We note that since the volume field is constructed from the halos, there is a residual correlation analogous to the standard shot noise in the halo auto-power spectrum. We measure the cross-power spectrum between the density field and the volume field obtained from the random catalog with the same number density as the halo field and subtract it from  $ P_{ \rm h \mathcal{V} } $. Both the results before and after this shot-noise subtraction are shown. We indeed find that the shot-noise subtracted results are in good agreement with the prediction $ b_{\rm h m} b_{ \mathcal{V}  \rm m} $, for which we have used the fit results from $ b_{\rm h m}$ and $ b_{ \mathcal{V}  \rm m}$ obtained with Eq.~\eqref{eq:b_fit}.

\subsection{The BAO in the volume field}

\begin{figure*}%[!htb]
\centering
\includegraphics[width=0.77\linewidth]{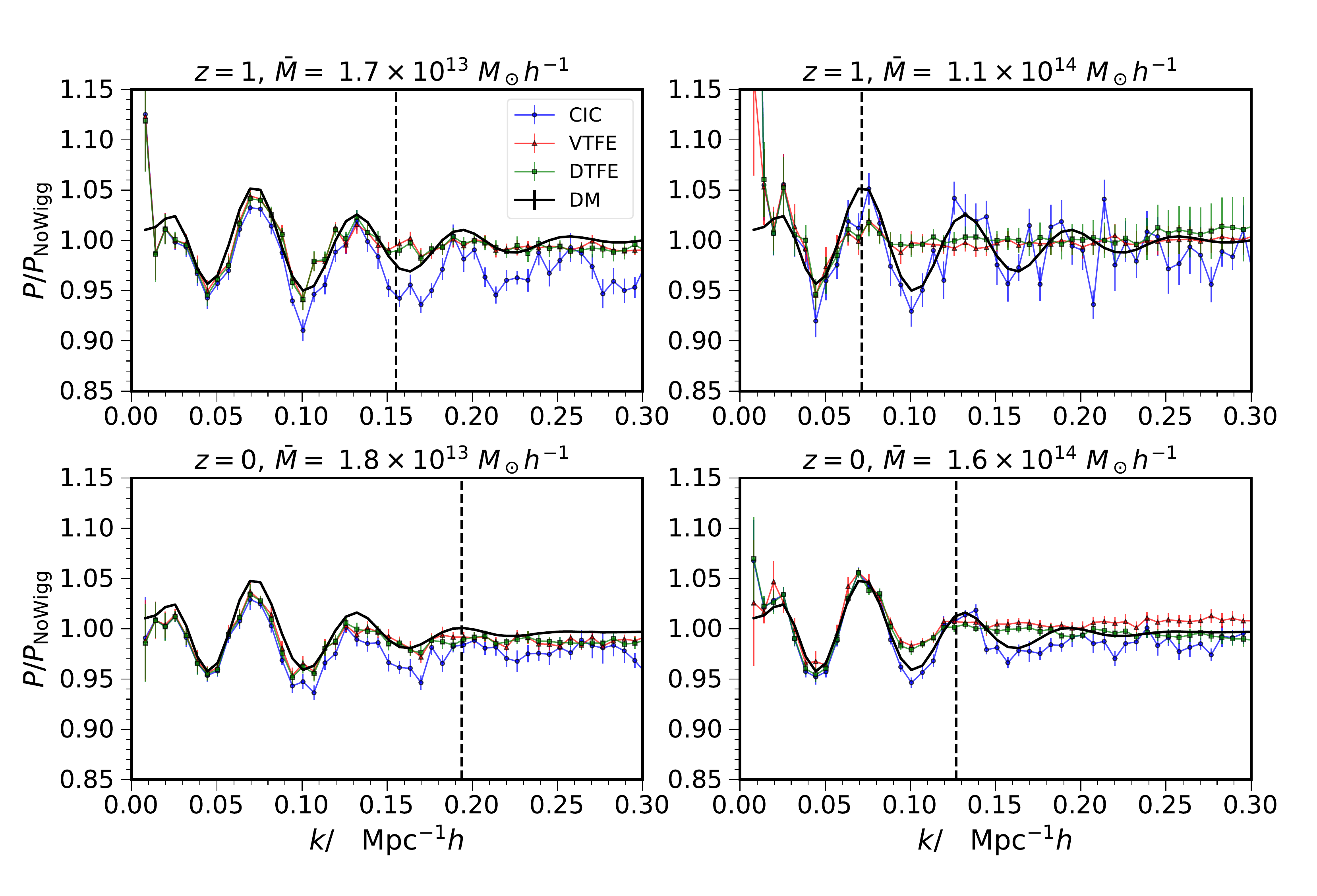}
\caption{Ratio between real-space halo density power spectra from simulations with fiducial and no-wiggle initial conditions. The results for different halo density fields obtained with the CIC (blue), VTFE (red), and DTFE (green) method are shown. The corresponding measurements from the dark matter (black solid curve) are overplotted as reference. The vertical dashed line indicates the mean separation scale $k_{\rm ms}$ of the halos. The halo density power spectra from the tessellation methods can reproduce the BAO features on large scales, while the power beyond $ k_{\rm ms} $ is suppressed.  } 
\label{fig:BAOratio_Pkdelta_real_subset}
\end{figure*}

\begin{figure*}%[!htb]
\centering
\includegraphics[width=0.77\linewidth]{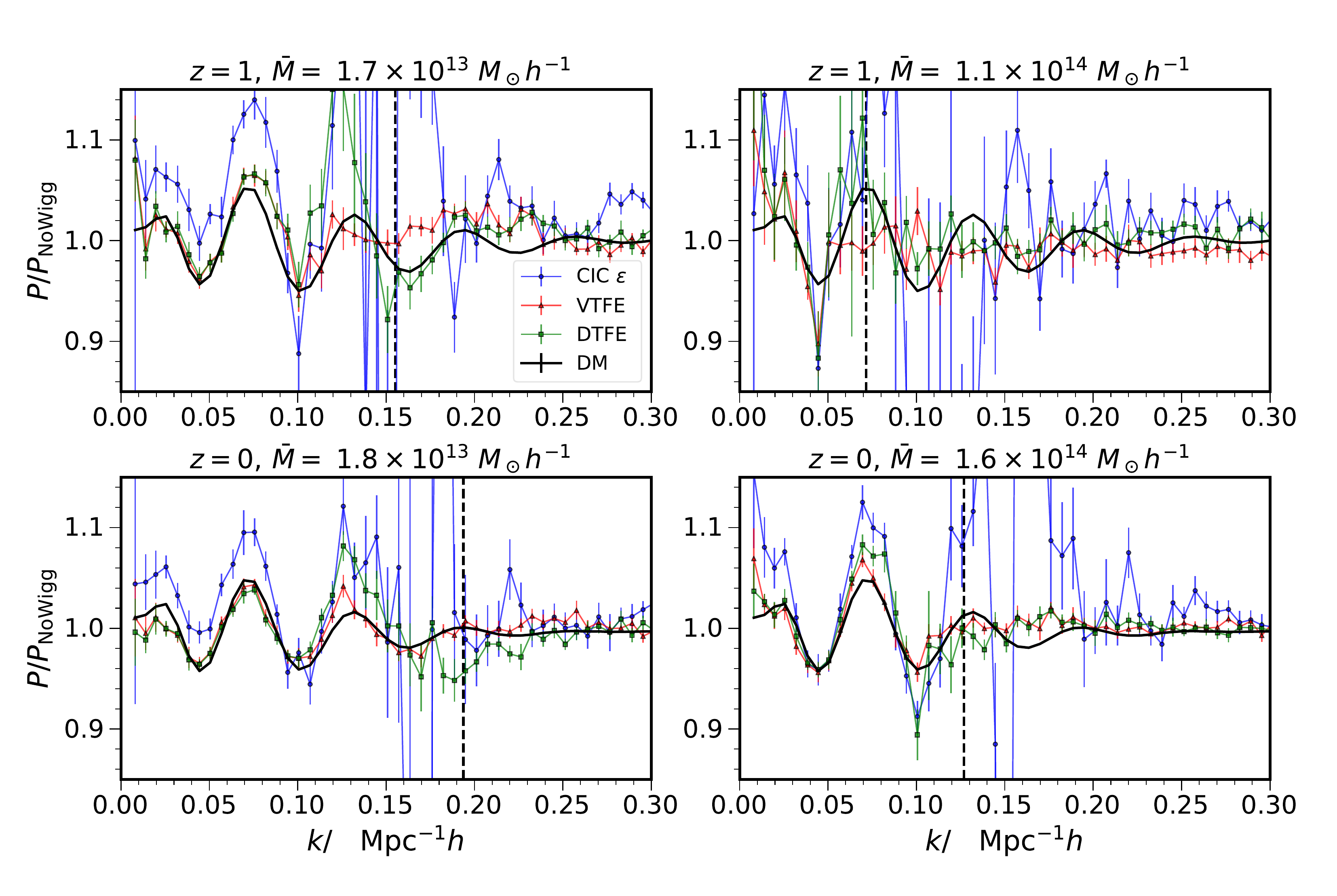}
\caption{As Fig.~\ref{fig:BAOratio_Pkdelta_real_subset}, but for the volume field constructed from the halo distribution. The black curve shows the BAO from the dark matter density field. The blue data points are the CIC measurement of the modified $ \mathcal{V}_\epsilon $ statistic with $\epsilon  = 0.1 $. On large scales, the volume statistics computed with the tessellation methods exhibit BAO features without systematic bias. }
\label{fig:BAOratio_PkVol_real_subset}
\end{figure*}

%\clearpage

In the early Universe, photons and baryons couple to form a hot plasma in which acoustic oscillations are excited.  These oscillations leave important imprints in large-scale structure of the late Universe \cite{PeeblesYu1970,SunyaevZeldovich1970}. The BAO features are regarded as one of the most important probes of the large-scale structure  and have been detected  in numerous galaxy surveys \cite{WeinbergMortonson_etal2013}.

Physically, the BAO manifests itself as an excess probability of finding galaxies at a distance  $r_{\rm d}$, the sound horizon at the drag epoch. This appears as a peak at the scale $r_{\rm d} $ in the galaxy density correlation function and as oscillations (wiggles) in Fourier space. Analogously,  given a depression of $\mathcal{V} $ at some location, it is more likely to find another depression of $\mathcal{V} $ at a distance of $r_{\rm d}$. This also gives rise to a positive enhancement in the correlation function of $\mathcal{V} $  at the scale of $r_{\rm d}$. We now go on to investigate the anticipated BAO signals in the volume statistic in more detail. To do so we use two sets of simulations, one with the fiducial setup, and another one with the Eisenstein-Hu initial conditions without BAO wiggles~\cite{EisensteinHu_1998}.

\subsubsection{Real space}

We begin with the BAO features measured in the real-space density field using the tessellation methods. To highlight the BAO features, we show the ratio between power spectra from the fiducial and the no-wiggle initial conditions in Fig.~\ref{fig:BAOratio_Pkdelta_real_subset}. The results for different halo groups obtained with three interpolation methods are compared. The BAO feature in the dark matter field is also shown for reference, which is determined with the CIC method. The number density of dark matter particles is so high that its density field can be regarded as continuous here. 

On large scales (small $k$) all methods produce similar results, as the large-scale modes are unaffected by the interpolation methods. At higher $k$ (compared to $k_{\rm ms}$), the CIC halo field is still able to reproduce the BAO features imprinted in the dark matter density field, albeit with more noise. On the other hand, for the VTFE and DTFE fields the BAO wiggles start to be smoothed out close to $ k_{\rm ms} $ and are suppressed significantly at $ k \gtrsim  k_{\rm ms} $.  Following Appendix~\ref{sec:window_Pk}, when the particle distribution is smoothed by a constant window function, the window factors out and cancels in the power spectrum ratio.  Thus, the fact that there is smoothing of the ratio between the wiggle and no-wiggle power spectrum implies that the smoothing effect cannot be attributed to a constant window and it  must arise from the density-dependent nature of the tessellation methods.  Numerous works have shown that the nonlinearity of density fields causes a smoothing of the BAO and its effect on the power spectrum can be approximated by a Gaussian window \cite{Bharadwaj1996ZA,EisensteinSeoWhite2007,Crocce:2007dt,Matsubara:2008wx}. This damping of the BAO is primarily driven by the large-scale bulk flow motion.  Because the tessellation methods adaptively track the evolution of the particle distribution, the associated window has a similar effect, resulting in a smoothing of the BAO wiggles.  This interpretation on the tessellation window is consistent with results in Sec.~\ref{sec:tessellation_as_effectivewindow}.

%Thus, the smoothing of the  BAO wiggles must arise from the density-dependent nature of the tessellation methods. Numerous works have shown that the nonlinearity of density fields causes a smoothing of the BAO and its effect on the power spectrum can be approximated by a Gaussian window \cite{Bharadwaj1996ZA,EisensteinSeoWhite2007,Crocce:2007dt,Matsubara:2008wx}. Because the tessellation methods adaptively track the evolution of the particle distribution, the associated window has a similar effect, resulting in a smoothing of the BAO wiggles.  

We now turn to the BAO imprints in the volume field, as presented in Fig.~\ref{fig:BAOratio_PkVol_real_subset}. The BAO measurement in the volume field is noisier than that in the density. In particular, although the VTFE and DTFE behave similarly for the density case, the DTFE yields more noisy results than the VTFE for the volume field. Overall, the BAO wiggles imprinted in the volume field follow those in the matter density field without any systematic bias. As in the density case above, beyond  $ k_{\rm ms } $ the BAO signature is washed out.

In the standard CIC interpolation, $\mathcal{V} $ is ill-defined for regions with $ \delta_{\rm h} = -1 $, which happens for empty grid cells. To overcome this problem, we can instead define
\begin{align}
\mathcal{J}_\epsilon  & = \frac{ 1 }{ 1 + \delta_{\rm h} + \epsilon }, \\ \nn
\mathcal{V}_\epsilon  & =  \frac{\mathcal{J}_\epsilon  }{ \bar{\mathcal{J}}_\epsilon } - 1, 
\end{align}
with some constant $\epsilon>0$, which ensures that $ \mathcal{V}_\epsilon $ is always well-defined. We show the results for $\epsilon = 0.1$ in Fig.~\ref{fig:BAOratio_PkVol_real_subset}. This method yields noisier results than the tessellation methods. We have checked that other values of $\epsilon$, such as 0.01 or 0.2, do not improve this.

\begin{figure*}%[!htb]
\centering
\includegraphics[width=0.78\linewidth]{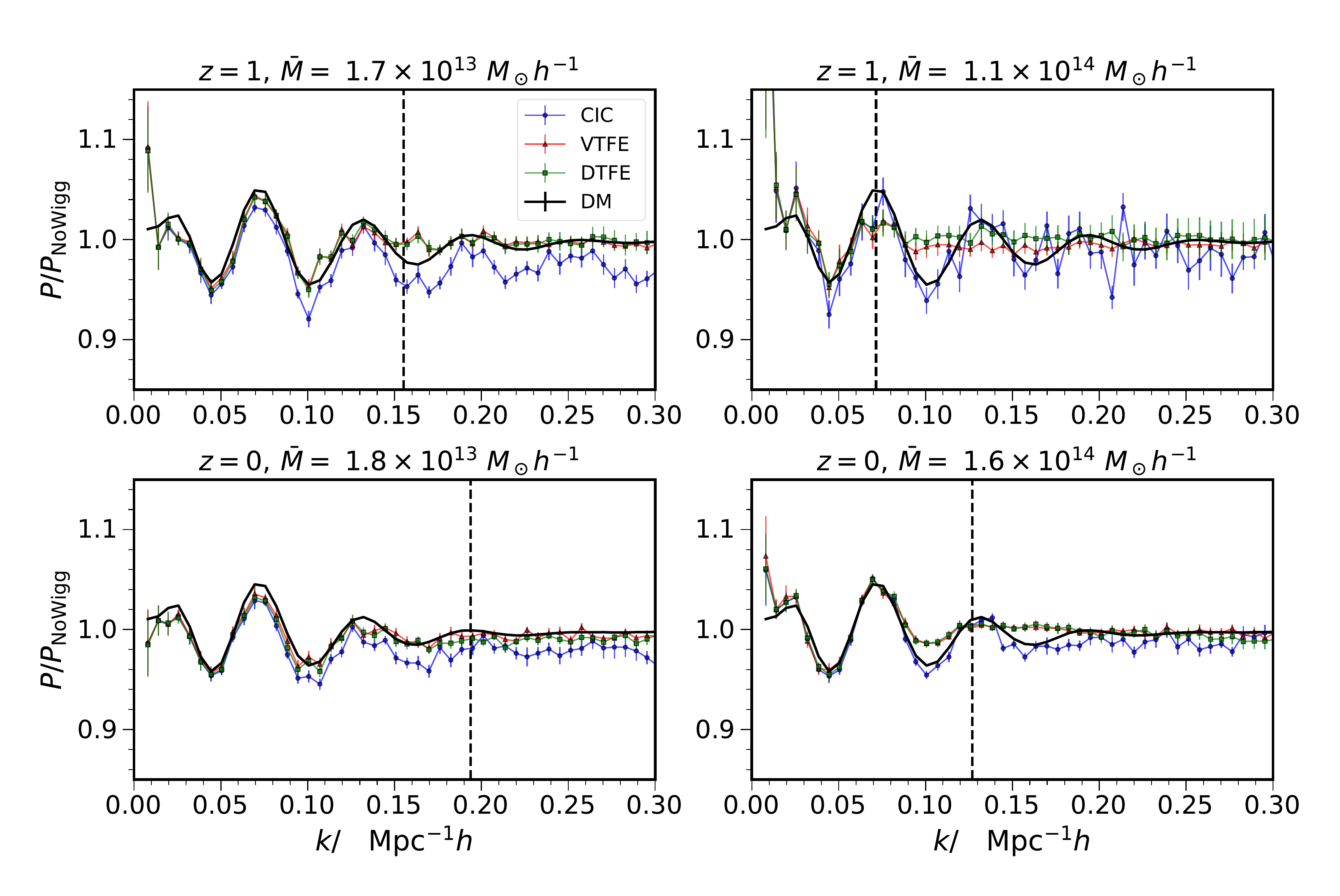}
\caption{As Fig.~\ref{fig:BAOratio_Pkdelta_real_subset}, but for the monopole power spectrum of the density statistics in redshift space.  Similar to the results in real space, apart from the smoothing for $k\gtrsim k_{\rm ms} $, the tessellation methods accurately reproduce the BAO wiggles in redshift space.  }
\label{fig:BAOratio_Pkdelta_ell0_subset}
\end{figure*}

\begin{figure*}%[!htb]
\centering
\includegraphics[width=0.78\linewidth]{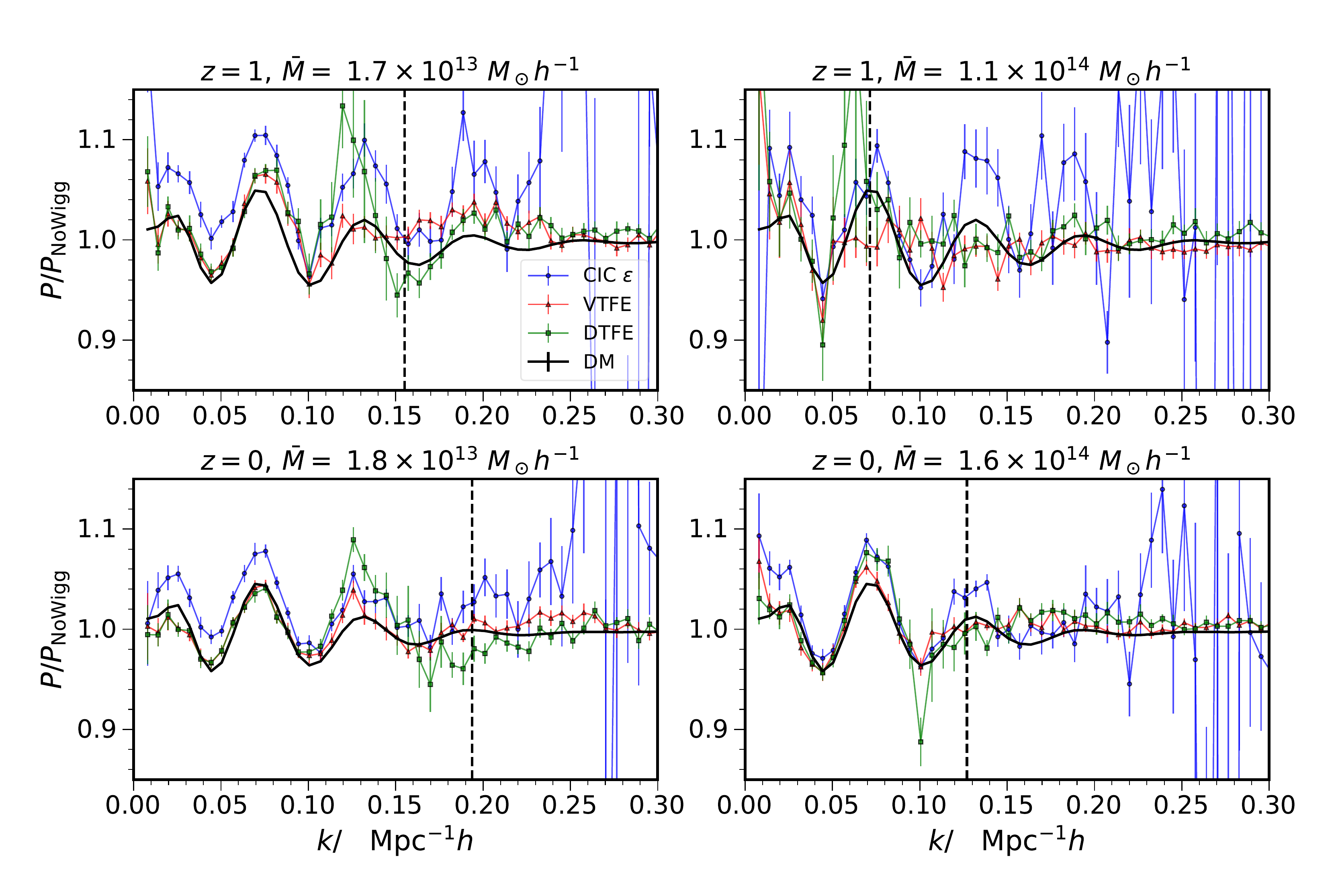}
\caption{As Fig.~\ref{fig:BAOratio_PkVol_real_subset}, but for monopole power spectrum of the volume statistics in redshift space. Similar to the results in real space, the large-scale BAO features are imprinted in the redshift-space volume power spectra without systematic bias. }
\label{fig:BAOratio_PkVol_ell0_subset}
\end{figure*}

%\clearpage

\subsubsection{Redshift space}

Observationally, we can only perform BAO measurements in redshift space. As the coordinates of the galaxies along the line-of-sight direction are deduced from redshifts in galaxy surveys, the density field is subject to additional perturbations due to their peculiar motion. These perturbations cause RSD, and we shall consider them in the plane-parallel limit, with the comoving coordinate in the $z$-direction $ x_z $ modified to 
\beq
\label{eq:RSD_transformation}
s_z = x_z + \frac{v_z }{ a H }
\eeq
where $ v_z$ is the peculiar velocity in $z$-direction  and $H$ is the Hubble parameter. We can adapt Eq.~\eqref{eq:delta_h_continuity} to redshift space as
\beq
( 1 + \delta_{\rm L} ) d^3 q = \big( 1 + \delta_{\rm h}( \bm{x}) \big) \Big| \frac{ \partial \bm{x} }{ \partial \bm{s} } \Big|d^3 s = \big( 1 + \delta_{\rm h}(\bm{s}) \big) d^3 s.  
\eeq
Accordingly, we have 
\beq
\mathcal{J} = \frac{ d^3 s }{ ( 1 + \delta_{\rm L} )  d^3 q }   = \frac{1}{ 1 + \delta_{\rm h} (\bm{s}) }. 
\eeq
Hence, the extension of the volume statistic to redshift space is straightforward.

The redshift-space power spectrum can be expressed in terms of multipoles,
\beq
P_\ell = \frac{2 \ell + 1 }{2} \int_{-1}^{1} d \mu P_s(k,\mu)  \mathcal{L}_\ell(\mu),
\eeq
where $ P_s $ is the power spectrum in redshift space, $\mu$ the cosine of the angle between $\hat{\bm{k}}$ and the line of sight, and $ \mathcal{L}_\ell $ is the Legendre polynomial of order $\ell$.  Here we only show results for the monopole of the halo power spectrum, as the quadrupole measurements are noisy even for 20 realizations.

The monopole power spectrum for the halo density field in redshift space is shown in Fig.~\ref{fig:BAOratio_Pkdelta_ell0_subset}. The results are similar to the real-space case. The halo monopole from the CIC interpolation appears to trace the wiggles in the dark matter monopole power spectrum well. While the tessellation results are slightly less noisy than the CIC ones for $ k< k_{\rm ms} $, they are suppressed for  $ k \gtrsim  k_{\rm ms} $.  Fig.~\ref{fig:BAOratio_PkVol_ell0_subset} displays the monopole power spectrum for the volume statistics. As in real space, $\mathcal{V}_\epsilon $ yields the most noisy estimator and the VTFE results are the most robust in reproducing the dark matter BAO.  We note that the BAO amplitude in the volume field appears slightly enhanced at some scales. This is particularly apparent for the $\mathcal{V}_\epsilon $  estimator, which is most sensitive to noise. We therefore attribute this effect to the discreteness of the tracer distribution.

%We note that the signals for the volume field is mildly stronger in redshift space than that in real space. This is particularly apparent for the $\mathcal{V}_\epsilon $ curve, in which the noisy part in Fig.~\ref{fig:BAOratio_PkVol_real_subset}  appears at  higher $k$ in Fig.~\ref{fig:BAOratio_PkVol_ell0_subset}.     

\subsubsection{Discussion}

Besides the auto-power spectrum, we can measure the BAO using the cross-power spectrum between $\delta_{\rm h} $ and $\mathcal{V}$. The results are similar to those obtained from the auto-power spectrum on large scales, but they are more noisy for $ k \gtrsim k_{\rm ms} $, since the volume field lacks the BAO feature on small scales. So far we have exclusively investigated Fourier-space statistics. In configuration space, the effect of the tessellation is a smoothing of the BAO peak as well. Thus, the BAO feature measured from the correlation function of the volume field is broadened and becomes less sharp, making it harder to differentiate from the broad-band correlation function.

In order to fully capture the BAO features in the power spectrum of the volume statistics, the number density of the tracer sample must be sufficiently high. For example, at $z=0$ the halo sample with mean mass $1.8 \times 10^{13} \Msun  $ and number density $2.3 \times 10^{ -4 } \, (\MpcOh)^{-3} $ is sufficient. At higher redshift, the BAO wiggles are less damped by nonlinearities, so a higher number density is necessary to push $ k_{\rm ms} $ to a larger value. For instance, at $z=1$ the halo number density needs to be at least $7 \times 10^{ -4 } \, (\MpcOh)^{-3} $ to fully capture the BAO wiggles.

To summarize the virtues of each method: for the density statistics the CIC interpolation method is recommended, as it offers an unbiased estimate of the small-scale density field, while the VTFE and DTFE smooth out the field for scales above $k_{\rm ms}$. The tessellation imposes additional conditions such that the resultant field is smooth and space-filling. These requirements modify the small-scale behavior of the field. On the other hand, in order to exploit volume statistics the tessellation method is preferred, as it is able to construct a space-filling field with non-vanishing density everywhere.   The results from the VTFE are similar to the DTFE for the density statistics, but VTFE yields more robust results for the volume statistics. Another advantage of the VTFE is its lower computational overhead compared to DTFE\footnote{In a typical run with a single core on an Intel Xeon E5-2686 cluster, the VTFE takes about 40 minutes, while the DTFE about 24 hours. Perhaps the DTFE algorithm can be more efficient after further optimization.}.   Hence, for the clustering analysis of volume statistics, we recommend usage of the VTFE.

A measurement of the BAO feature from the distribution of underdense regions is interesting on its own, but it may also provide valuable information on cosmology in addition to what is available from halo clustering alone. We have demonstrated that the volume statistic traces the large-scale structure with a negative bias parameter. What remains to be shown is how correlated the volume statistic and the traditional halo density statistic are. Since the volume statistic is constructed via the halo tracer distribution, the answer is not obvious. For example, one can construct a trivial field $-\delta_{\rm h } $, which exhibits negative linear bias\footnote{However, for voids the behavior of quadratic bias is similar to that of halos \cite{Chan:2014qka,Chan:2019yzq}, so at second order the bias of this artificial field is opposite to that of the genuine underdense tracer.}, but perfectly correlates with $\delta_{\rm h} $. In order to investigate the correlation between $\delta_{\rm h}  $ and $\mathcal{V}  $, one could determine the covariance of the power spectrum from both the density and the volume field using many different realizations, which is beyond the scope of this paper.

A simple (albeit less conclusive) test is to consider the cross-correlation coefficient between  $\delta_{\rm h} $ and $\mathcal{V}$
\beq
r = \frac{ P_{\rm h \mathcal{V}}  }{\sqrt{  P_{\rm hh }  P_{ \mathcal{VV}} } },
\eeq
where $ P_{\rm hh }$, $ P_{ \mathcal{VV}} $, and  $ P_{\rm h \mathcal{V}} $ are the halo auto-, volume auto-, and halo-volume cross-power spectra.

\begin{figure*}%[!htb]
	\centering
	\includegraphics[width=0.8\linewidth]{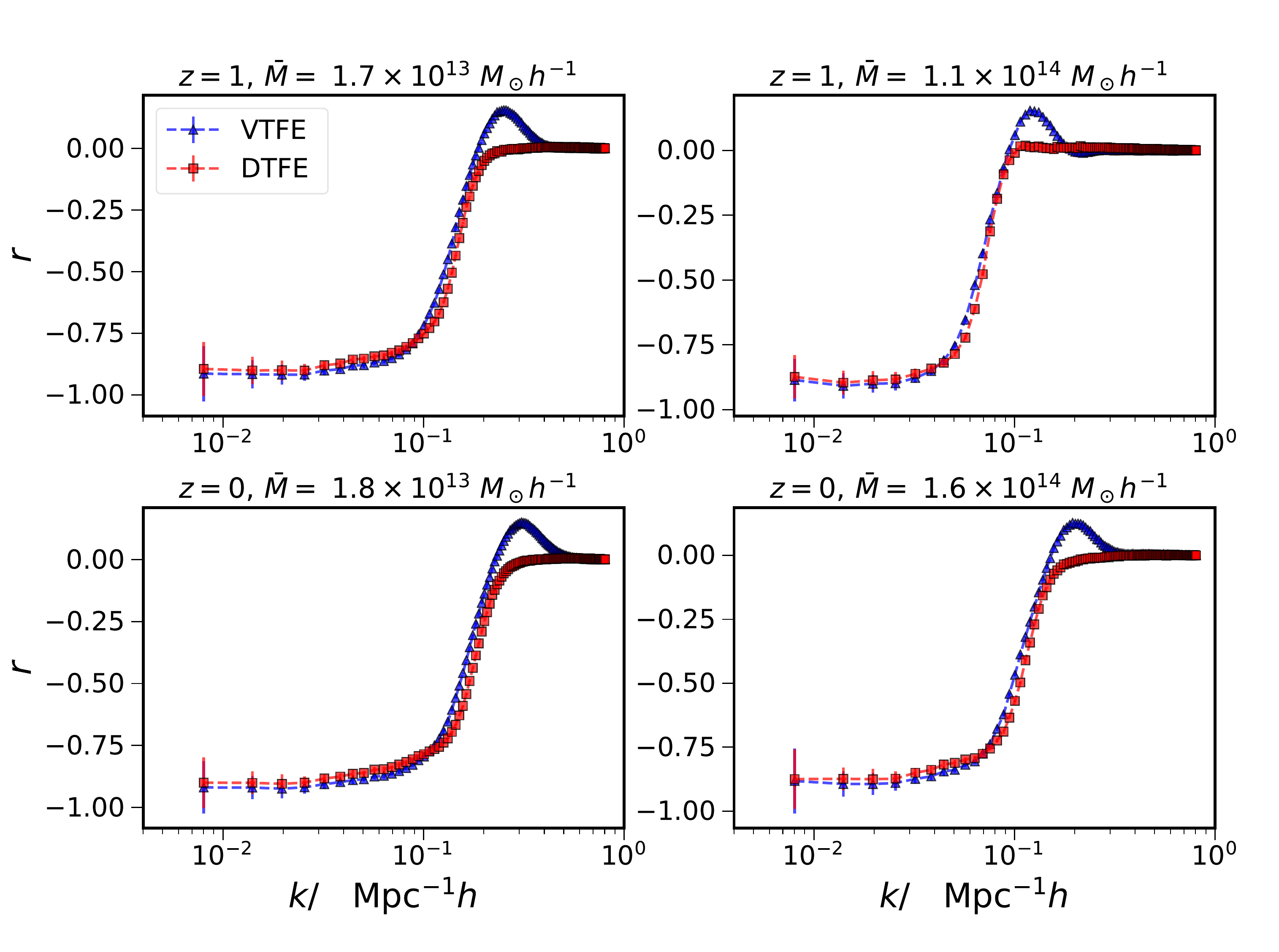}
	\caption{Cross-correlation coefficient between the halo density field $\delta_{\rm h}$ (estimated via CIC) and the volume field $\mathcal{V}$. The latter is estimated using the distribution halos as tracers, either via VTFE (blue triangles), or via DTFE (red squares).}
	\label{fig:r_hv_correlationcoef_redshift}
\end{figure*}

The results for fields constructed from four different halo groups are shown in Fig.~\ref{fig:r_hv_correlationcoef_redshift}. We have used the full redshift-space monopole power spectra including shot noise. On large scales, both  $\delta_{\rm h}  $ (estimated via CIC) and $\mathcal{V}  $ (estimated via VTFE or DTFE) are (anti-) correlated with the dark matter $\delta$ and with each other, but due to the presence of shot noise and other sources of stochastic noise in the volume field, $r$ is slightly above $-1$.   The characteristic shape of the curve is due to the exclusion between $\delta_{\rm h} $ and $\mathcal{V} $.   At smaller scales, exclusion can lead to oscillations before $r$ approaches to zero, but they are quickly suppressed when the shot noise kicks in at high $k$.   If the volume statistics were as trivial as $ - \delta_{\rm h} $, $r $ would be equal to $ -1 $ on all scales.

We note that the BAO features are also detectable using overlapping voids \cite{Liang:2015oqc,Kitaura:2015ubm}, which are defined to be the circumspheres of the tetrahedrons resulting from the Delaunay tessellation of the galaxies \cite{Zhao:2015ecx}. Although this approach is similar to ours in using the tessellation to partition the point set, we do not define objects on it. There can be substantial overlap between the circumspheres of neighboring tetrahedrons. In the void construction algorithm, the void center and size are determined entirely by the four particles spanning the tetrahedron. The point-like nature of the void-center position prevents any additional smoothing of the BAO; however, the resultant BAO measurement may be more correlated with the density statistic.

Finally we comment on the possible effects of the BAO reconstruction \cite{EisensteinSeoWhite2007}. BAO reconstruction is often applied in galaxy surveys (e.g.~\cite{Padmanabhan_etal2012}) to undo part of the large-scale gravitational evolution and RSD effect so as to boost the BAO signal.  This is because the reconstructed field becomes more correlated with the initial conditions, so the BAO signal in the volume field is expected to increase as well. For example,  $\mathcal{V}$  reduces to  $-\delta_{\rm h} $  for weak fluctuations and we expect the BAO signal to be well correlated between these two fields. However, a detailed study is required to access the overall gain in BAO information from the reconstructed volume field.

%On one hand, the reconstructed field is more correlated with the initial conditions and  the BAO signal in the volume field is expected to increase as well. On the other hand, for weaker fluctuations, $\mathcal{V}$ would reduce to $-\delta_{\rm h} $ to a good approximation, and we expect the signal in the volume field to be more correlated with that in the density field. Thus a detailed study is required to access the overall gain in the BAO information from the reconstructed field. 

\section{Conclusions}
\label{sec:conclusions}

In large-scale structure analyses it is common to exploit statistics based on the galaxy distribution, which predominantly traces high-density regions in the Universe.  However, a large fraction of its volume exhibits relatively low density. Hence, the clustering of underdense regions may bear cosmologically relevant information that is complementary to the conventional clustering of overdensities. In principle, cosmic voids are ideal proxies for probing the underdense regions, but their auto-correlation statistics suffer a large shot-noise contamination due to their low number density. In this work we propose to apply the volume statistic $\mathcal{V}$ to probe the volume distribution of large-scale structure. This statistic provides a measure of the volume change between Eulerian and Lagrangian space. As it makes use of all the tracer particles available, its shot-noise level is similar to that of the conventional tracer density field. Furthermore, the definition of the volume statistic is closely related to the density contrast, so it may be more amenable to theoretical models based on perturbation theory than objects that are defined via nonlinear topological characteristics of the cosmic web, such as voids. Also, an extension of its definition to redshift space is straightforward.

Traditional mass-assignment methods, such as CIC interpolation, yield empty regions with $\delta_{\rm h} =-1$ for sparse halo samples, which make the volume statistic estimator ill-defined there. To overcome this difficulty, we apply tessellation interpolation methods to estimate the volume field. We study the clustering statistics obtained via the tessellation methods in detail. The power spectrum from the tessellated field is smoothed by an effective window function, which on large scales can be approximated by the same simple function for different tessellation methods. On small scales this approximation fails due to the density dependence of the tessellation.

For the density clustering, the conventional mass interpolation methods are recommended because they can reproduce the original discrete point distribution for sufficiently large grid size. Because of the additional smoothness condition imposed, the tessellation methods cause smoothing of the field at scales smaller than the mean particle separation.  When the tessellation is used for the volume statistics, the VTFE provides the most satisfactory results in terms of signal-to-noise ratio. At the same time the VTFE is computationally less expensive than the DTFE.

The clustering amplitude of the volume statistic is negatively biased with respect to the dark matter density field $\delta$ on large scales, reflecting the fact that it traces underdense regions in the Universe. This bias can also be measured without knowledge of $\delta$ via the observable cross-power spectrum between the fields $ \delta_{\rm h}  $ and $\mathcal{V}$, which are strongly anti-correlated on large scales. Furthermore, the BAO features are imprinted in the volume statistic as well. Apart from a smoothing beyond the mean tracer separation scale $k \gtrsim k_{\rm ms} $, they are reproduced without systematic bias. Hence, to avoid loss of any clustering information, the number density of the tracer used to generate the volume statistic must be sufficiently high. An investigation of the covariance between the BAO information in the density and volume statistic is left for future work.

As the signal-to-noise ratio of void clustering is significantly limited by shot noise, we expect volume statistics to complement the available information from the clustering of underdense regions.   For example, a cross-correlation with galaxy positions and shapes, quasars, the CMB and even three-point statistics of the volume statistic should be detectable at high significance. Voids have been suggested to be a sensitive probe for rich phenomena, such as the imprints of dark energy, modified gravity and massive neutrinos \cite{Pisani:2019cvo}. The volume statistic opens up a new avenue towards exploring these ideas. For example, in analogy to Ref.~\cite{Chan:2018piq} the bias of the volume statistic can be used to constrain the inflationary paradigm via PNG. This can be combined with the multi-tracer approach~\cite{Seljak:2008xr, McDonald:2008sh, HamausSeljakDesjacques2011, Hamaus:2012ap, Abramo:2013awa} to optimize the statistical gain of this method.

\section*{Acknowledgments}
We thank Xin Wang and Yi Zheng for useful discussions.  The simulations in this work were run at the Kunlun cluster at the School of Physics and Astronomy of the Sun-Yat Sen University.  K.C.C. acknowledges the support from the National Science Foundation of China under the grant 11873102 and the Science and Technology Program of Guangzhou, China (No. 202002030360). N.H. is supported by the Excellence Cluster ORIGINS, which is funded by the Deutsche Forschungsgemeinschaft (DFG, German Research Foundation) under Germany's Excellence Strategy -- EXC-2094 -- 390783311.

\appendix

\section{Kernel Density Estimation (KDE)}
\label{sec:Density_KDE}
In this Appendix, we show that the density can be estimated using the method of Kernel Density Estimation (KDE) in statistics. See e.g.~\cite{LiRacine_KDE} for more details on KDE.  Given a number of data points $\bm{x}_\alpha $ sampling the underlying probability distribution, instead of estimating the probability density by histogram, the sample points are smoothed by a window (or kernel) $W$ of characteristic size $h_\alpha$. Note that the stochastic variables are the sampling points and the underlying probability density is fixed.  The probability density $p$ can be estimated by the estimator 
\beq
\hat{p}(\bm{x} ) \Delta V = \frac{1}{N} \sum_{\alpha=1   }^N W \left(  \frac{\bm{x}- \bm{x}_\alpha }{ h_\alpha  }  \right),
\eeq
where $N$ is the total number of particles contributing to density estimation. For overlapping windows it includes all the particles available. The volume element $ \Delta V $ is to be fixed later on.
The expectation of $ \hat{p}(\bm{x} ) $ is
\begin{align}
  \langle \hat{p}(\bm{x} ) \rangle &= \frac{1 }{N  \Delta V} \sum_{\alpha=1}^N \left\langle W \left( \frac{\bm{x} - \bm{x}_\alpha }{ h_\alpha }  \right) \right\rangle \nn \\
  & = \frac{1 }{N  \Delta V } \sum_{\alpha=1}^N  \int d^3 x_\alpha   W\left( \frac{\bm{x} - \bm{x}_\alpha }{ h_\alpha }  \right) p ( \bm{x}_\alpha ) \nn\\
  & = \frac{1 }{N  \Delta V } \sum_{\alpha=1}^N  \int d^3 y_\alpha h_\alpha^3   W \left( \bm{y}_\alpha  \right) p ( \bm{x} - h_\alpha  \bm{y}_\alpha ). 
\end{align}
By expanding the probability density to second order we have
\begin{align}
  \langle \hat{p}(\bm{x} ) \rangle &\approx   \frac{1 }{N  \Delta V } \sum_{\alpha=1}^N  \int d^3 y_\alpha h_\alpha^3   W \left( \bm{y}_\alpha  \right) \Big[  p ( \bm{x})  \nn \\
    &- h_\alpha \sum_i \bm{y}_{\alpha i} \frac{\partial p}{\partial \bm{x}_i } + \frac{1}{2} \sum_{ij}  h_\alpha^2  \bm{y}_{\alpha i} \bm{y}_{\alpha j}  \frac{\partial^2 p}{\partial \bm{x}_i \bm{x}_j } \Big]  . 
\end{align}
The window satisfies the property that
\beq
\int d^3 y W(\bm{y}) = 1
\eeq
and it is chosen to be an even function. Therefore it simplifies to
\begin{align}
  \langle \hat{p}(\bm{x} ) \rangle & =   \frac{ \sum_{\alpha=1}^N  h_\alpha^3  }{N  \Delta V }  p ( \bm{x}) + \frac{ \sum_{\alpha =1}^N h_{\alpha}^5  }{2N \Delta V  }   \sum_{ij} \frac{\partial^2 p }{ \partial x_i \partial x_j  }  I^{(1)}_{ij}    ,   
\end{align}
where for convenience we define the notation
\beq
\label{eq:I_integral} 
I_{ ij\dots k}^{(n)} \equiv  \int d^3y \, W^n(y) y_i y_j \dots y_k . 
\eeq
For $\hat{p} $ to be unbiased to the lowest order, we require
\beq
 \Delta V =   \frac{1 }{N} \sum_{\alpha=1}^N  h_\alpha^3  . 
\eeq
Then the bias arises from the second derivative term:
\begin{align}
  \label{eq:KDE_bias}
 \langle \hat{p}(\bm{x} ) \rangle  -  p(\bm{x}) =   \frac{  \sum_{\alpha =1}^N h_{\alpha}^5  }{2  \sum_{\alpha =1}^N h_{\alpha}^3 }  \sum_{ij} \frac{\partial^2 p }{ \partial x_i \partial x_j  }  I^{(1)}_{ij}    .    
\end{align}
In particular, at the peak (trough), the estimated probability density is lower (higher) than the true one.

If we assume that the sampling points are independent of each other, the variance of $ \hat{p} $ is given by
\begin{align}
  \mathrm{Var}\left(   \hat{p}(\bm{x})  \right) &= \frac{1 }{ ( N \Delta V )^2 } \sum_{\alpha =1 }^N   \mathrm{Var} \left(  W \Big(\frac{\bm{x} - \bm{x}_\alpha }{h_\alpha }  \Big) \right)  \nn \\
  & =  \frac{1 }{ ( N \Delta V )^2 } \sum_{\alpha =1}^N  \Big\{  \int dx_\alpha W^2\Big( \frac{\bm{x} - \bm{x}_\alpha   }{ h_\alpha  } \Big) p(x_\alpha ) \nn \\
  & -   \Big[ \int dx_\alpha W\Big( \frac{\bm{x} - \bm{x}_\alpha   }{ h_\alpha  } \Big) p(x_\alpha) \Big]^2       \Big\}.
\end{align}
To the lowest order in $h$ we have   
\begin{align}
    \label{eq:KDE_variance}
  \mathrm{Var}\left(   \hat{p}(\bm{x})  \right) & \approx \frac{I^{(2)} p(\bm{x})   }{\sum_{\alpha =1 }^N  h_\alpha^3 }.
\end{align}
Note that the variance is inversely proportional to the total volume of the windows.

Now suppose that we have $N$ particles, and we would like to estimate the number density of the sample, $ n(\bm{x})$, by applying a smoothing window to the particles.  This can be done by simply replacing $p$ by $ n(\bm{x}) / N $, which is non-negative and normalized to 1.  The estimator for $ n$ reads
\begin{align}
  \label{eq:n_KDEestimator} 
\hat{n}( \bm{x} )  = \frac{N}{ \sum_{\alpha=1}^N  h_\alpha^3   } \sum_{\alpha=1   }^N W \left(  \frac{\bm{x}- \bm{x}_\alpha }{ h_\alpha  }  \right).
\end{align}
The bias and the variance of the estimator are given by
\begin{align}
  \label{eq:KDEn_bias}
  \langle  \hat{n} (\bm{x} )  \rangle - n(\bm{x}) &= \frac{\sum_{\alpha =1}^N  h_\alpha^5 }{ 2  \sum_{\alpha =1}^N  h_\alpha^3 } \sum_{ij} \frac{\partial^2 n  }{\partial x_i \partial x_j }  I_{ij}^{(1)},  \\
  \label{eq:KDEn_var}
  \mathrm{Var}( \hat{n} (\bm{x} ) ) &= \frac{ n(\bm{x}) I^{(2)}  }{ \frac{1}{N} \sum_{\alpha=1}^N h_\alpha^3  }  .
\end{align}
These results apply to the density estimation using the usual mass interpolation method. It is important to remember that these results are obtained assuming that the sampling points Poisson sample the underlying density field.

\section{Effects of the window function on the power spectrum}
\label{sec:window_Pk}
The standard shot noise power spectrum result [Eq.~\eqref{eq:Pshot}] can be derived by considering the particle distribution given by a sum of Dirac deltas; see e.g.,~the Appendix of \cite{Chan:2016ehg}.  Using the number density given by 
\beq
n(\bm{x}) = \sum_\alpha \Ddel( \bm{x} - \bm{x}_\alpha ), 
\eeq
the discrete two-point correlation can be written as 
\begin{align}
  \label{eq:ShotPoisson}
   & \langle n(\bm{x}_1 )   n(\bm{x}_2 )\rangle \nn \\
 = &  \Ddel( \bm{x}_1 - \bm{x}_2 )  \Big\langle   \sum_\alpha \Ddel( \bm{x}_1 - \bm{x}_\alpha ) \Big\rangle \nn + \bar{n}^2 [ 1 + \xi(x_{12} ) ]   \\
 = &  \bar{n}  \Ddel( \bm{x}_1 - \bm{x}_2 ) +   \bar{n}^2 [ 1 + \xi(x_{12} ) ]  ,
\end{align}
where $ \bar{n} $ is the mean number density given by
\beq
\bar{n} = \big\langle   \sum_\alpha \Ddel( \bm{x} - \bm{x}_\alpha )   \big\rangle, 
\eeq
and $ \xi $ is the correlation function due to continuous clustering. The discrete correlation function then reads
\beq
\xi_{\rm d} ( \bm{x}_1 - \bm{x}_2 )  =  \frac{1}{ \bar{n} }  \Ddel( \bm{x}_1 - \bm{x}_2 ) +  \xi(x_{12} ). 
\eeq
Upon Fourier transform, we obtain
\beq
P_{\rm d} ( k ) = \frac{1 }{(2 \pi)^3 \bar{n}  } + P(k). 
\eeq
with the first term being Eq.~\eqref{eq:Pshot}.

The Dirac delta function can be thought of as a special kind of window function.  We now consider the case when the particles are smoothed by an extended window function $W$. The number density then reads 
\beq
n_W(\bm{x}) = \sum_\alpha W( \bm{x} - \bm{x}_\alpha ). 
\eeq
Note that because  the window smoothing respects mass conservation, $ \bar{n} =  \langle n_W \rangle $. 
Then the two-point correlation reads
\begin{align}
     & \langle n_W(\bm{x}_1 ) n_W(\bm{x}_2 ) \rangle   \nn \\
  = & \bar{n}^2   \int d^3 x'  \int d^3 x''     W( \bm{x}_1 - \bm{x}' )  W( \bm{x}_2 - \bm{x}'' )  \xi(|  \bm{x}' -  \bm{x}'' |)  \nn \\
& +   \bar{n}^2  +   \bar{n}  \int d^3 x' W( \bm{x}_1 - \bm{x}' ) W( \bm{x}_2 - \bm{x}' ) . 
\end{align}
The discrete correlation can then be written as 
\begin{align}
 &  \xi_{WW}( | \bm{x}_1 - \bm{x}_2 | )   = \frac{1}{ \bar{n} }  \int d^3 x' W( \bm{x}_1 - \bm{x}' ) W( \bm{x}_2 - \bm{x}' ) \nn \\
 &   \quad  +   \int d^3 x'  \int d^3 x''     W( \bm{x}_1 - \bm{x}' )  W( \bm{x}_2 - \bm{x}'' )  \xi(|  \bm{x}' -  \bm{x}'' |)  
\end{align}
In Fourier space we have
\begin{align}
P_{WW}( k ) = (2 \pi )^6 |W(k)|^2  \left[ P(k)  +  \frac{1}{ (2 \pi)^3 \bar{n} } \right]. 
\end{align}
The first term in the square  bracket is analogous to the 2-halo term and the second is the analog of the 1-halo term in the halo model \cite{Cooray:2002dia}. For both the continuous and the shot noise term, the effect of the window function is the same and can be accounted for by a $ |W(k)|^2  $ factor.

Similarly the cross-correlation function between $n$ and $n_W $ is given by
\begin{align}
 &  \xi_{W}( | \bm{x}_1 - \bm{x}_2 | )   = \frac{1}{ \bar{n} }   W( \bm{x}_1 - \bm{x}_2 )  \nn \\
 &   \quad  +   \int d^3 x'  \int d^3 x''     W( \bm{x}_1 - \bm{x}' )  \Ddel( \bm{x}_2 - \bm{x}'' )  \xi(|  \bm{x}' -  \bm{x}'' |)  
\end{align}
and the cross-power spectrum reads
\begin{align}
 P_W( k ) = (2 \pi)^3 W(k)  \left[ P(k)  +  \frac{1}{ (2 \pi)^3 \bar{n} } \right]. 
\end{align}
Again we find that the effect of the window function is captured by a factor of  $W(k) $.

\bibliography{reference} 

\end{document}